\documentclass[notitlepage,prd,superscriptaddress,nofootinbib]{revtex4}
 \usepackage{color}
\usepackage{amsmath}
\usepackage{amssymb}
\usepackage{amscd}
\usepackage{latexsym}

\def\be{\begin{equation}}
\def\ee{\end{equation}}
\def\arr{\begin{array}{rll}}
\def\ea{\end{array}}
\def\bea{\begin{eqnarray}}
\def\eea{\end{eqnarray}}

\def\N2{$N{=}2$}

\def\sfrac#1#2{{\textstyle\frac{#1}{#2}}}
\def\>{\rangle}
\def\<{\langle}
\def\+{\dagger}
\def\={\ =\ }

\newcommand{\ben}{\begin{enumerate}}
\newcommand{\een}{\end{enumerate}}
\newcommand{\beq}{\begin{equation}}
\newcommand{\eeq}{\end{equation}}
\newcommand{\bse}{\begin{subequation}}
\newcommand{\ese}{\end{subequation}}
\newcommand{\bc}{\begin{center}}
\newcommand{\ec}{\end{center}}

\newcommand{\nn}{\nonumber \\}

\newcommand{\ch}{{\tt h}}

\def\DC{\kern2pt {\hbox{\sqi I}}\kern-4.2pt\rm C}
\def\DH{\rm I\kern-1.5pt\rm H\kern-1.5pt\rm I}

\renewcommand{\=}{\ =\ }

\newcommand{\bQ}{{\overline Q}}

\begin{document}
\title{Symmetries of deformed
supersymmetric mechanics on K\"ahler manifolds }
\author{Evgeny Ivanov}
\email{eivanov@theor.jinr.ru}
\affiliation{Bogoliubov Laboratory of Theoretical Physics, Joint Institute for Nuclear Research, Dubna, Russia}
\author{Armen Nersessian}
\email{arnerses@yerphi.am}
\affiliation{Yerevan Physics Institute, 2 Alikhanian Brothers St., Yerevan  0036 Armenia}
\affiliation{Yerevan State University, 1 Alex Manoogian St., Yerevan  0025 Armenia}
\affiliation{Bogoliubov Laboratory of Theoretical Physics, Joint Institute for Nuclear Research, Dubna, Russia}
\author{Stepan Sidorov}
\email{sidorovstepan88@gmail.com}
\affiliation{Bogoliubov Laboratory of Theoretical Physics, Joint Institute for Nuclear Research, Dubna, Russia}
\author{Hovhannes Shmavonyan}
\email{hovhannes.shmavonyan@yerphi.am}
\affiliation{Yerevan Physics Institute, 2 Alikhanian Brothers St., Yerevan 0036 Armenia}

\begin{abstract}
Based on the systematic  Hamiltonian and superfield approaches  we construct the  deformed $\mathcal{N}=4,8$  supersymmetric mechanics on K\"ahler manifolds
 interacting with constant magnetic field, and study their symmetries.
At first we  construct the deformed $\mathcal{N}=4,8$ supersymmetric Landau problem  via minimal coupling of standard (undeformed) $\mathcal{N}=4,8$ supersymmetric
 free particle systems on K\"ahler manifold with   constant magnetic field.
We show that the initial ``flat'' supersymmetries are necessarily deformed to
 $SU(2|1)$ and $SU(4|1)$ supersymmetries, with the magnetic field playing the role of deformation parameter, and  that the resulting systems
 inherit all the kinematical symmetries of the initial ones.
 Then we  construct $SU(2|1)$ supersymmetric K\"ahler oscillators and find that they include,
 as particular cases, the  harmonic oscillator models on  complex Euclidian and complex projective  spaces, as well as superintegrable deformations thereof, viz.
 $\mathbb{C}^N$-Smorodinsky-Winternitz and $\mathbb{CP}^N$-Rosochatius systems.
 We show that the supersymmetric extensions  proposed  inherit all the  kinematical symmetries of the initial bosonic models.
  They also inherit, at least in the case of  $\mathbb{C}^N$ systems, hidden (non-kinematical) symmetries.
 The superfield formulation of these  supersymmetric  systems is presented, based on the worldline $SU(2|1)$ and $SU(4|1)$ superspace formalisms.

\end{abstract}
\maketitle
\setcounter{page}{1}
\setcounter{equation}0
\section{Introduction}
The models of supersymmetric  mechanics were initially introduced as  toy models  for  supersymmetric field theories. However, it was shortly realized that
 such models are of big interest  on their own right. An important feature of the supersymmetric mechanics models is that the main new ingredient they bring in, the fermionic variables,
after quantization become the operators representing the spin of particle. As the result, the fermionic parts of the relevant Hamiltonians play the role of generalized Pauli terms
describing an interaction of spin with external fields, in particular, with the magnetic field.
 From this viewpoint, the study of supersymmetric extensions of the mechanical systems interacting  with the magnetic field is of obvious importance.
 However, such systems seem not to have attracted enough attention,  despite an enormous number of publications on supersymmetric mechanics.

This is rather surprising, having in mind that  the first practical  application of (${\mathcal N}=2$) supersymmetric mechanics technique was
the explanation of the ``accidental''  double degeneracy of the spectrum of  the (planar) Landau problem (see, e.g.,~\cite{GK}),
{\it i.e.}, the problem of  the planar motion of non-relativistic electron (charged $\frac12$-spin  particle) in a constant magnetic field.
For a long time it has been one of the central issues treated in the textbooks on quantum mechanics~\cite{LL}.
However, nowadays, saying  ``Landau problem'', people sometimes ignore  the spin of the original system.

The compact (spherical) analog  of the planar Landau
problem is associated with a particle moving on the two-sphere in the presence of constant magnetic field generated by a Dirac monopole placed in the center
of the sphere. The spherical Landau problem enjoys ${SO}(3)$ invariance which is also characteristic of the ``free'' particle on the two-sphere. The higher-dimensional generalization of
this problem,  a particle on $\mathbb{CP}^N$ interacting with a constant magnetic field, inherits $SU(N+1)$ invariance of the relevant free system.
Quantum mechanically, the inclusion of constant magnetic field  supplies the system with a degenerate ground state,
 which is just due to the preservation of the symmetries of a free particle.
Thanks to this degeneracy, the quantum-mechanical Landau problem constitutes the  basis  of the  theory of quantum Hall effect \cite{hall}, equally as of its higher-dimensional
generalizations to complex projective spaces \cite{cpn}.

It is more or less obvious that the inclusion of constant fields preserves the initial symmetries of the free particle moving on the generic K\"ahler manifold as well,
and  (spinless) Landau problem can be defined for any K\"ahler manifold. In order to restore the initial  meaning of the  Landau problem in the context of these systems
one  should try to  construct supersymmetric extensions of the (spinless) Landau problem on K\"ahler manifold, such that they preserve the initial kinematical symmetries.
However, in the existing literature devoted to supersymmetric extensions of the (generalized) Landau problem, the discussion of symmetry properties of the supersymmetric systems constructed
 is as a rule  left aside (see, e.g., \cite{hasebe,nlin}).\\

 While for $\mathcal{N}=2$ the construction of such supersymmetric extensions is a rather trivial task,
 it is not the case for $\mathcal{N}\geq 4$  supersymmetric extensions. Generically, one may pose the question:\\

How should systems on K\"ahler manifolds in interaction with a constant magnetic fields (in particular,  the Landau problem) be supersymmetrized, so that their initial symmetries
be preserved?\\

We guess that  the general answer is as follows. Instead of considering $\mathcal{N}, d=1$ Poincar\'e supersymmetric extensions of given bosonic systems,
one should  deal with superextensions based on the proper deformations of standard $d=1$ Poincar\'e supersymmetry.

An attempt towards proving this conjecture was performed years ago in \cite{CPosc}, where  it was observed that the oscillator and the Landau problem on  a complex projective space
admit the deformed $\mathcal{N}=4$  supersymmetric extension (later on called  ``weak $\mathcal{N}=4$  supersymmetric extension'' \cite{smilga}),
which preserves the initial kinematical symmetries of those systems. Departing from this model, the class of systems  with non-zero potentials called ``K\"ahler oscillator''
was introduced \cite{CPosc,Kahlerosc}, such that they admit similar deformed supersymmetric extensions respecting the inclusion of constant magnetic field. The relevant bosonic
Hamiltonian reads
\be\label{Kahosc}
{H}_{osc}=g^{\bar a b}\left({\bar \pi}_a\pi_b +|\omega|^2\partial_{\bar{a}}K\partial_b K \;\right),
\ee
where $K(z,\bar z)$ is the K\"ahler potential.

A few years later, the  one-dimensional version of  that K\"ahler superoscillator model
  was re-derived  within a $d=1$ superfield formalism   based on $SU(2|1)$ superalgebra
 treated as a deformation of $\mathcal{N}=4, d=1$ Poincar\'e superalgebra \cite{ivanovsidorov,ISKahler}.
 Thereby, the  ``weak ${\cal N}=4$  supersymmetry'' was identified with $su(2|1)$ superalgebra (this fact was also independently noticed in the paper \cite{kor}
treating supersymmetric quantum Landau problem on  $\mathbb{CP}^1$ ). Using similar techniques, the deformed  $\mathcal{N}=8$ one-dimensional Landau  problem
associated with $su(4|1)$ superalgebra was also defined \cite{ILS19rev}.
This study was to large extent inspired by the activity on building field-theoretical  models with  the ``rigid supersymmetry on curved superspaces'' initiated in
\cite{sca}.

Having in mind the "practical importance" of supersymmetrization   respecting symmetries of initial system
 and field-theoretical importance of "curved superspace approach",  we present here the systematic   approach to the  deformed supersymmetrization of the systems on K\"ahler manifolds interacting  with  the constant magnetic field

Having in mind the ``practical importance'' of supersymmetrization   respecting symmetries of the initial bosonic system
 and the field-theoretical importance of the ``curved superspace approach'',  we  develop here the systematic
 approach to the  deformed supersymmetrization of systems ``living'' on K\"ahler manifolds and interacting  with  a constant magnetic field
 by the use of a {\sl  supersymmetric analog of a minimal coupling}.
In the superfield formulations, such a coupling naturally comes out under some minimal choice of the related superfield Lagrangians.

Resorting first to the Hamiltonian formalism,  we construct in this way the $SU(2|1)$ supersymmetric  extensions of the K\"ahler oscillator (and of the Landau problem) on the generic K\"ahler space, as well as
the $SU(4|1)$ supersymmetric Landau problem on the special K\"ahler manifolds of the rigid type (that is a K\"ahler manifold
equipped with the holomorphic symmetric tensor of the third rank obeying some compatibility condition \cite{fre}).
We show that this approach perfectly matches with the requirement that the supersymmetric Landau problem exhibits all the kinematical symmetries of the
original system and involves the appropriate spin interaction. It is demonstrated that both $SU(2|1)$ and $SU(4|1)$
supersymmetric Landau problems inherit all the kinematical symmetries of the initial systems. Requiring the
Hamiltonian in the $SU(2|1)$ case to commute with all supercharges  amounts to adding the appropriate Zeeman term to it. In the superspace language, this means that
we should start from the properly central-charge extended superalgebra, with the Hamiltonian being identified with the relevant central charge.
Analogously, the general $SU(2|1)$ K\"ahler superoscillator systems as superextensions of those with the Hamiltonian \eqref{Kahosc} can
be constructed and then reproduced from the superfield approach.

Exemplifying the general analysis, we set up and study $SU(2|1)$ supersymmetric extensions of the following  particular superintegrable  K\"ahler oscillator models:
\begin{itemize}
\item $\mathbb{C}^N$-oscillator (the sum of $N$ two-dimensional isotropic oscillators);
\item  $\mathbb{C}^N$-Smorodinsky-Winternitz system (the sum of $N$ copies of two-dimensional isotropic oscillators deformed by  ring-shaped potentials)\cite{shmavonyan};
 \item $\mathbb{CP}^N$-oscillator \cite{CPosc,quantCPosc},  i.e. the  $\mathbb{CP}^N$- counterpart of  $\mathbb{C}^N$-oscillator;
\item $\mathbb{CP}^N$-Rosochatius system \cite{Ros}, i.e.  the $\mathbb{CP}^N$- counterpart of  $\mathbb{C}^N$-Smorodinsky-Winternitz system.
\end{itemize}
We show that these models also inherit all the kinematical symmetries of the initial systems. In addition, we find the explicit expressions for the superanalogs  of the
hidden symmetry generators  of  $\mathbb{C}^N$-oscillator and $\mathbb{C}^N$-Smorodinsky-Winternitz system ({\it i.e.}, of the Fradkin and Uhlenbeck tensors).
Unfortunately, we were not yet able to find the superanalogs of such hidden symmetry generators for the $\mathbb{CP}^N$-oscillator  and of the $\mathbb{CP}^N$-Rosochatius system,
though they hopefully exist. \\

The paper is organized as follows:\\

In {\sl Section 2} we describe the phase superspace as a proper setting for supersymmetrization of systems on K\"ahler manifolds in interaction with a constant magnetic field.
The Legendre transformation relating Hamiltonian and Lagrangian formulations of those systems is given.
In {\sl Section 3} we present the   Hamiltonian  formulations of $SU(2|1)$ and $SU(4|1)$ supersymmetric Landau problems. The general Hamitonian formulation of $SU(2|1)$
K\"ahler superoscillator is described in {\sl Section 4}. As an example, we show that this class of Hamiltonians incorporates the supersymmetric version of two-dimensional
anisotropic oscillator. In {\sl Section 5} the previously considered systems are recovered within the manifestly $SU(2|1)$ and $SU(4|1)$ covariant off-shell superfield approaches.
{\sl Section 6} is devoted to a more detailed discussion of the  $SU(2|1)$ supersymmetric extensions of the oscillator-like systems on $\mathbb{C}^N$ and $\mathbb{CP}^N$
that are listed above and to the study of their symmetries.

\setcounter{equation}0
\section{Phase superspace, kinematical symmetries, and Lagrangians}
The K\"ahler manifold $M$ is the Hermitian manifold  with the Hermitian metrics, $ds^2=g_{a\bar b} dz^a d\bar z^b$, which
also defines the symplectic structure
 \be
 \omega_M=i g_{a\bar b}dz^a\wedge d\bar z^b,\; d\omega_M=0\quad \Rightarrow\quad g_{a\bar b} =
\partial_a \partial_{\bar{b}}\,K \,, \qquad \partial_a = \frac{\partial}{\partial z^a}\,, \; \partial_{\bar{b}}
  =\frac{\partial}{\partial{\bar z}^{b}}\,,
 \ee
  where the real function $K(z,\bar z)$,
 K\"ahler potential, is defined up to holomorphic and antiholomorphic functions: $K(z,\bar z)\to K(z,\bar z)+U(z)+{\bar U}(\bar z)$.

The K\"ahler manifold  can be equipped with the  Poisson brackets associated with the above symplectic structure
\be
\{f,g\}_M=i g^{\bar a b}\Big(\frac{\partial f}{\partial {\bar z}^a} \frac{\partial g}{\partial {z}^b}-\frac{\partial g}{\partial {\bar z}^a} \frac{\partial f}{\partial{z}^b}\Big), \quad g^{\bar a{ b}}g_{\bar{b}c}=\delta^a_c\; .
\ee
Therefore, the  isometries of  K\"ahler structure should preserve both complex and symplectic structures, {\it i.e.}, they are generated
by  the holomorphic  Hamiltonian vector fields,
 \begin{equation}
{\bf V}_{\mu}=\{\ch_\mu, \}_M =
    V_\mu^{a}(z)\frac
{\partial}{\partial z^a}+
{V}_\mu^{\bar a}(\bar z) \frac{\partial}
{\partial \bar z^a}, \quad  V^a_\mu=i g^{\bar{b}a}\partial_{\bar{b}}\ch_\mu(z,\bar z) \;,\quad  V^{\bar a}_\mu =\overline{V^{a}_\mu}\,,
\label{iso}\end{equation}
where the real function $\ch_\mu (z,\bar z)$   is a momentum map sometimes  called Killing potential.
The holomorphicity  of the vector field yields the following equation to the Killing potential
\begin{equation}
\frac{\partial^2 \ch_\mu}{\partial z^a \partial z^b} -
\Gamma^c_{ab}\frac{\partial \ch_\mu}{\partial z^c}=0
,
\label{v}\end{equation}
 with $\Gamma^c_{ab} = g^{c\bar{d}}\partial_a g_{b \bar{d}}\,$
 \footnote{The only non-vanishing components of the Christoffel symbol in the K\"ahler geometry
are $\Gamma^c_{ab}$ and $\Gamma^{\bar c}_{\bar{a}\bar{b}} = g^{d\bar{c}}\partial_{\bar{a}} g_{d \bar{b}}$.}.
The same result can be obtained  by the direct solving of  the Killing equations
\be
({\rm a})\;V_{\mu a;b}+V_{\mu b;a}=0,\qquad ({\rm b})\;V_{\mu a;\bar b}+V_{ \mu \bar b;a}=0,\quad{\rm  with}\quad V_{\mu a}=g_{a\bar b}V_{ \mu }^{\bar b}. \label{Killing}
\ee
The  action of the vector field $\mathbf{V}_{\mu}$ on an arbitrary
function $f(z,\bar z)$ can be expressed through
the Poisson bracket with the Killing potential
$$
\mathbf{V}_{\mu}\,f = \{\ch_\mu, f\}_M\,.
$$
Thus, the requirement that the vector fields $\mathbf{V}_{\mu}$ form Lie algebra  amounts to  the same Lie algebra relations for Killing potentials
\be
[{\bf V}_{\mu},{\bf V}_{\nu}]=C_{\mu \nu}^{\lambda}{\bf V}_{\lambda} ,\quad \Leftrightarrow \quad \{ \ch_{\mu},\ch_{\nu}\}_M=C_{\mu \nu}^{\lambda} \ch_{\lambda}+ {\rm const},
\ee
where the constant term either corresponds to co-circle in that Lie algebra or can be absorbed by the appropriate constant shift of Killing potentials.\\

Let us consider the electrically charged particle moving on a K\"ahler manifold and interacting  with the constant magnetic field of  strength $B$, i.e. the $U(1)$-Landau problem on K\"ahler manifold.
 For this aim   we
equip  the cotangent bundle  of the K\"ahler  manifold
with the following  symplectic structure and Hamiltonian
\begin{equation}
\omega_B=
d\pi_a\wedge dz^a +
d{\bar\pi}_{a}\wedge d{\bar z}^{a} -i B g_{a\bar b}dz^a\wedge
d{\bar z}^b,\qquad { H}_0= g^{\bar a b}{\bar \pi}_a\pi_b  .
\label{ssB}\end{equation}
The corresponding Poisson brackets are given by
\be
\{\pi_a, z^b\}=\delta^b_a,
 \quad
\{\pi_a,\bar\pi_b\}= i B\,g_{a\bar b}. \label{BosPB}
\ee
The isometries of a K\"ahler
 structure discussed earlier define the
Noether constants of motion
 \begin{equation}
{J}_\mu=V^{a}_\mu\pi_a +
 {\bar V}^{\bar a}_\mu {\bar\pi}_{\bar a} -B\ch_\mu( z\bar z ), \quad  V^a_\mu=i g^{\bar{b}a}\partial_{\bar{b}}\ch_\mu(z,\bar z) \; :\quad
\left\{
\begin{array}{c}
  \{{ H}_0, J_{\mu}\}_B=0 \\
  \{J_\mu, J_\nu\}_B=C_{\mu\nu}^\lambda J_\lambda
\end{array}\right\},
\label{jmu}\end{equation}
where the brackets $\{\cdot\,, \cdot\}_B$ are calculated according to \eqref{BosPB}.
Notice that the vector fields
 generated by ${ J}_\mu$ are independent of $B$,
\begin{equation}
{\bf{\tilde V}_{\mu}}=\{J_{\mu},\quad\}_{B}=V_{\mu}^a(z)\frac{\partial}{\partial z^a}-V^a_{\mu,b}\pi_a\frac{\partial}{\partial \pi_b}\;  + \quad{\rm c.c.}\; .
\label{mommap}\end{equation}
Hence, coupling to a
 constant magnetic field preserves
the whole symmetry algebra of a free
particle moving on a K\"ahler manifold.  This implies that the Landau problem can be properly defined on any  K\"ahler manifold.

To construct fermionic extensions  of the systems on K\"ahler manifolds interacting with constant magnetic field
we define the  $(2N|MN)_{C}$-dimensional  phase superspace
equipped with the symplectic structure
\begin{equation}
\begin{array}{c}
\Omega=d\pi_a\wedge dz^a+ d{\bar\pi}_a\wedge d{\bar z}^a
-i(B g_{a\bar b}-R_{a{\bar b}c\bar d}\eta^{c\alpha}\bar\eta^d_\alpha)
dz^a\wedge d{\bar z}^b+
 i g_{a\bar b}D\eta^{a\alpha}\wedge{D{\bar\eta}^b_\alpha}\quad,
\end{array}
\label{ss}\end{equation}
where $\alpha=1,\ldots M$  are spinorial indices, $D\eta^{a\alpha}
=d\eta^{a\alpha}+\Gamma^a_{bc}\eta^{b\alpha} dz^c$, and
 $\Gamma^a_{bc},\; R_{a \bar b c\bar d}=g_{e\bar b}(\Gamma^e_{ac})_{,\bar d}$ are,
respectively, the components of connection and curvature of
the K\"ahler structure.

The Poisson brackets  corresponding to the symplectic structure \eqref{ss} amount to the
relations
\be
\{\pi_a, z^b\}=\delta^b_a,\quad
\{\pi_a,\eta^{b\alpha}\}=-\Gamma^b_{ac}\eta^{c\alpha},\quad
\{\pi_a,\bar\pi_b\}=i(Bg_{a\bar b}- R_{a\bar b c\bar d}\eta^{c\alpha}{\bar\eta}^d_\alpha),
\quad
\{\eta^{a\alpha}, \bar\eta^b_\beta\}=
i g^{a\bar b}\delta^{\alpha}_{\beta}
\label{spbobr}\ee
and their complex conjugates. They induce the following generic Poisson bracket for the functions on the phase superspace
\be
\{f,g\}=\frac{\partial f}{\partial\pi_a}\wedge\nabla_a g +\frac{\partial f}{\partial\bar\pi_a} \wedge{\bar\nabla}_a g +i(Bg_{a\bar b}-R_{a\bar b c\bar d}\eta^{c\alpha}{\bar\eta}^d_\alpha)\frac{\partial f}{\partial\pi_a}\wedge\frac{\partial g}{\partial\bar\pi_b}+i g^{\bar a b}\left(\frac{\partial^l f}{\partial\eta^{a\alpha}}\wedge\frac{\partial^r g}{\partial\bar\eta^b_\alpha}\right),
\label{sPB}\ee
where $A\wedge B= AB-(-1)^{p(A)p(B)}BA $ and
\be
\nabla_a\equiv \frac{\partial}{\partial z^a}-\Gamma^c_{ab}\eta^{b\alpha}\frac{\partial}{\partial \eta^{c\alpha}}.
\ee
The extended symplectic structure \eqref{ss} and Poisson brackets \eqref{sPB}
are manifestly covariant with respect to the transformation
\be
{\widetilde z}^a={\widetilde z}^a(z), \qquad {\widetilde\pi}_a=\frac{\partial z^b}{\partial{\widetilde z}^a}\pi_b, \qquad{\widetilde\eta}^{a\alpha}=\frac{\partial{\widetilde z}^a}{\partial z^b}\eta^{b\alpha}.
\ee
Hence,
we can  lift the isometries \eqref{mommap} to the whole phase superspace  and define the respective super-Hamiltonian vector fields as
\be
\mathbf{V}_\mu=\{ \mathcal{J}_\mu, \;\}= V^a_\mu (z)\frac{\partial}{\partial z^a}-V^a_{\mu,b}\pi_a
\frac{\partial}{\partial \pi_b}+V^{a}_{\mu ,b}\eta^{b\alpha}\frac{\partial}{\partial \eta^{a\alpha}}\;+\;{\rm c.c.}\;,
 \label{sNeth0}\ee
 where
 \be
 {\cal J}_\mu={J}_\mu +\frac{\partial^2 {\ch}_\mu}{\partial z^c\partial
 {\bar z}^d}\eta^{c\alpha}\bar\eta^{d}_\alpha,
 \label{sNeth} \ee
 with $J_\mu$ defined by \eqref{jmu}.\\

Note  that the symplectic structure \eqref{ss} can be represented as a locally exact one-form,
 \be
\Omega=d\mathcal{A}\;\quad \mathcal{A}=\pi_adz^a+\bar\pi_a d\bar z^a+i\frac{B}{2}({\partial_a K} dz^a-{\partial_{\bar a} K} d\bar z^a)+\frac{i}{2} g_{a\bar b}(\eta^{a\alpha} D\bar \eta^b_\alpha+\bar\eta^b_\alpha D \eta^{a\alpha}).
 \ee
 Then, by the Hamiltonian
\be
 \mathcal{H}=g^{\bar a b}\bar\pi_a \pi_b +\mathcal{U}(z,\bar z,\eta,\bar\eta), \label{GenH}
 \ee
 where the potential term $\mathcal{U}(z,\bar z,\eta,\bar\eta)$
will be defined later for each specific system, we can immediately write down the first order-Lagrangian with the action
\be
\mathcal{S}=\int \mathcal{A}- \mathcal{H}dt.
\ee
Eliminating  cyclic variables $\pi_a, \bar\pi_a$, we  arrive  at the second-order Lagrangian
\be
\mathcal{L}= g_{a\bar b}\dot z^a\dot{\bar z}^b +
i\frac{B}{2}({\partial_a K} \dot z^a-{\partial_{\bar a} K} \dot{\bar z}^a)+\frac{i}{2} g_{a\bar b}(\eta^{a\alpha} D_t\bar \eta^b_\alpha+\bar\eta^b_\alpha D_t \eta^{a\alpha})
-\mathcal{U}(z,\bar z,\eta,\bar\eta)\;\quad {\rm with}\quad D_t\eta^a_\alpha
=\dot\eta^a_\alpha+\Gamma^a_{bc}\eta^b_\alpha \dot z^c.
\label{lagrgen}\ee

Now we can re-derive (and so check) all the previous formulas by applying the standard Legendre transformation just to this Lagrangian. We define
the canonical bosonic momenta
\be
P_a :=\frac{\partial {\cal L}}{\partial \dot{z}^a} = g_{a\bar b}{\dot{\bar z}}^b + i\frac{B}{2}\partial_a K - \frac{i}{2}\partial_c g_{a\bar{b}}\,(\eta^{c\alpha}\bar\eta^{\bar{b}}_\alpha)\,, \quad
P_{\bar a} :=\frac{\partial {\cal L}}{\partial \dot{\bar z}^a} = {\dot{ z}}^b g_{b\bar a}- i\frac{B}{2}\partial_{\bar a} K + \frac{i}{2}\partial_{\bar c} g_{b\bar{a}}\,
(\eta^{c\alpha}\bar\eta^{\bar{b}}_\alpha),  \label{CanP}
\ee
and the canonical fermionic ones
\be
P_{a\alpha}:=\frac{\partial^R \mathcal{L}}{\partial \dot{\eta}^{a\alpha}}\;=\frac{i}{2} g_{a\bar{b}} {\bar\eta}^{b}_\alpha,\quad
P_{\bar a}^\alpha :=\frac{\partial^R \mathcal{L}}{\partial \dot{\bar\eta}^{a}_\alpha}=\frac{i}{2}g_{\bar{a} b} \eta^{b \alpha}\;.
\ee
The above expressions indicate the appearance of    second-class constraints
\be
\phi_{a\alpha} = P_{a\alpha} - \frac{i}{2} g_{a\bar{b}} \bar\eta^{b}_\alpha \simeq 0\,, \quad \phi_{\bar a}^\alpha = P_{\bar a}^\alpha  - \frac{i}{2}g_{\bar{a} b} \eta^{b \alpha} \simeq
0\,. \label{2class}
\ee
Thus, for the Hamiltonian formulation we need to eliminate these constraints in accordance with  the Dirac's method.
The standard procedure  yields the following non-vanishing Dirac brackets (and their c.c.)
\bea
& \{ P_a,z^b  \} = \delta^b_a\,, \quad
 \{ P_a, \eta^{b\alpha} \} = -\frac12 \,\Gamma^b_{a c}\,\eta^{c\alpha}\,, \quad
\{ P_a, \bar\eta^{b}_{\alpha} \} = -\frac12 \, \partial_a g_{c \bar{d}}\, g^{c \bar{b}}\,\bar\eta^{d}_\alpha\,,\quad \{\eta^{a\beta}, \bar\eta^b_\alpha\}
= i g^{a\bar{b}}\delta^{\beta}_\alpha\,,& \nonumber\\
& \{ P_a\,,P_{\bar b}\} = -\frac{i}{4}\,\Big[\partial_a g_{c\bar{d}}\,\partial_{\bar{b}}  g_{f\bar{e}}
- (a\, \leftrightarrow \,\bar{b}) \Big] g^{c\bar{e}}\,(\eta^{f\alpha} \bar\eta^d_\alpha)\,, \quad \{ P_a\,,P_{b}\} =-
\frac{i}{4}\,\Big[\partial_a g_{c\bar{d}}\,\partial_{{b}}  g_{f\bar{e}}
- (a \,\leftrightarrow\, {b})\Big] g^{c\bar{e}}\,(\eta^{f\alpha} \bar\eta^d_\alpha).&
\label{Pbrack}
\eea
Introducing the non-canonical bosonic momenta $\pi_a=g_{a\bar b}\dot{\bar z}^b$, ${\bar\pi}_a={\dot z}^b g_{b\bar a}$ and taking into account
the relations between the momenta $P_a, \,P_{\bar b}, \, \pi_a, \, \pi_{\bar b}$ in \eqref{CanP} it is straightforward
to recover the brackets involving $\pi_a, \bar\pi_a$ and defined earlier in eqs. \eqref{spbobr}. In particular, it is easy to show that $\{\pi_a, \pi_b\} = \{\bar{\pi}_a,
\bar\pi_b\} = 0$. It is also straightforward, applying the Noether procedure
directly to \eqref{lagrgen} and assuming that the potential term $\mathcal{U}$ is invariant,  to reproduce the conserved isometry current ${\cal J}_\mu$ defined in \eqref{sNeth}.
With all these ingredients at hand, we are prepared to  turn to  supersymmetrizing the  Landau problem on K\"ahler manifold.

\section{Supersymmetric Landau problem}

To define the (deformed) ${\cal N}{=2M}$ supersymmetric extension of  Landau problem (i.e. of the  free particle interacting with a constant magnetic field)
 we make use of the strategy similar to symplectic coupling in the pure bosonic case.
The starting point is some
supersymmetric Hamiltonian system  supplied by supercharges
$Q^\alpha$ and~${\overline Q}_\alpha$
which close on a Hamiltonian~$\mathcal{ H}_0$,
\be
\{Q^\alpha, Q^\beta\}_0=\{{\overline Q}_\alpha, {\overline Q}_\beta\}_0=0\ ,\qquad
\{Q^\alpha, {\overline Q}_\beta\}_0=i \delta^{\alpha}_{\beta}\,{\cal H}_0\ ,\qquad
\{Q^\alpha, {\cal H}_0\}_0 = \{{\overline Q}_\alpha,{\cal H}_0\}_0 =0\ ,
\label{4susy}\ee
where the  Poisson brackets are defined by  \eqref{spbobr} with zero magnetic field,  $B=0$.

To introduce interaction with an external magnetic field, we deform
the supersymplectic structure, still preserving the form of the supercharges:
$
(\Omega_{B=0}\,, {Q}^\alpha, {\overline Q}_\alpha)\ \to\
(\Omega_B\,, {Q}^\alpha, {\overline Q}_\alpha)
$,
Now, the graded Poisson bracket~$\{{\cdot},{\cdot}\}$ is defined
through the symplectic form ~$\Omega_B$ defined in \eqref{ss},
and one has to check whether the supersymmetry algebra (\ref{4susy}) remains unaltered.

If this is the case, then
the Hamiltonian can be defined as $
{\cal H}_{0}\ :=\ \sfrac{i}{M}{\{Q^\alpha,{\overline Q}_\alpha\}}.
$
Otherwise we end up with some deformed  superalgebra which is different from the standard $d=1, \mathcal{N}=2M$ super Poincar\'e' algebra \eqref{4susy},
and we have to select there the generator admitting an interpretation as the appropriate Hamiltonian, i.e.
\be
\{Q^\alpha, Q^\beta\}=0+i B\ldots, \qquad \{Q^\alpha,{\overline Q}_\beta\}=i \delta^\alpha_{\beta}\mathcal{H}_{\rm 0}+i B\ldots
\ee
Here, dots  stand for some possible extra generators, which should be further commuted with supercharges and among themselves in order
to obtain a closed superalgebra.

Below we will  show that this program works perfectly well for the cases  of (deformed) $\mathcal{N}=4,8$ supersymmetric Landau problems.

\subsection{ The  $SU(2|1)$  (deformed  $\mathcal{N}=4$) supersymmetric Landau Problem}
In order to set up  $\mathcal{N}=4$ Landau problem we  choose the standard ``chiral" supercharges   $Q^\alpha, {\overline Q}_\alpha \,$ ($\alpha=1,2\,$)
with the same ansatz for them as in the absence of magnetic field, and introduce the charges generating the  $SU(2)$ $R$-symmetry
\be
Q^\alpha=\pi_a\eta^{a\alpha}\;,\quad {\overline Q}_\alpha=\bar\pi_a\bar\eta^a_\alpha, \quad\mathcal{R}^\alpha_\beta=g_{a\bar b}\eta^{a\alpha}\bar\eta^b_\beta - \frac{1}{2}\delta^\alpha_\beta g_{a\bar b}\eta^{a\gamma}\bar\eta^b_\gamma\;.
\label{TR}\ee
The closure of their Poisson brackets yields the superalgebra
\be
\begin{array}{c}
 \{Q^\alpha, Q^\beta\}=0, \quad   \{\mathcal{R}^\alpha_\beta,\mathcal{R}^\gamma_\delta\}=-i\delta^\gamma_\beta \mathcal{R}^\alpha_\delta + i \delta^\alpha_\delta\mathcal{R}^\gamma_\beta,\quad
 \{Q^\alpha, \mathcal{R}^\beta_\gamma\}= i\delta^{\alpha}_{\gamma}  Q^\beta - \frac{i}{2}\delta_{\gamma}^{\beta}  Q^{\alpha}\,,
\\[4mm]
\{Q^\alpha,{\overline Q}_\beta\}=i \delta^{\alpha}_{\beta}\mathcal{H}_{0} + i B \mathcal{R}^{\alpha}_{\beta},
\quad
\{Q^\alpha,\mathcal{H}_{0}\}=i \frac{B}{2} Q^\alpha,\quad\{\mathcal{R}^\alpha_\beta,\mathcal{H}_{0}\}=0\,,
\end{array}
\label{QR}\ee
where
 \be
\mathcal{H}_{0}=g^{\bar a b}\bar\pi_a\pi_b
-\frac12 R_{a\bar b c\bar d}\eta^{a\alpha}\bar\eta^b_\alpha\eta^{c\beta}\bar\eta^d_\beta + \frac B2
g_{a\bar b}\eta^{a\alpha}{\bar\eta}^b_\alpha . \label{hosup}
\ee
Extending the set \eqref{TR} by the  generator \eqref{hosup} we arrive at the $su(2|1)$ superalgebra (or ``weak $\mathcal{N}=4$ superalgebra'' in the terminology of \cite{smilga}).
We observe, however, that the  supercharges  do not commute with the Hamiltonian.
This drawback can be remedied via the appropriate modification of the Hamiltonian:
\be\label{hosup1}
\widetilde{\mathcal{H}}_{0}=\mathcal{H}_{0}-\frac{B}{2} g_{a\bar b}\eta^{a\alpha}\bar\eta^b_{\alpha}=g^{ a \bar b}\pi_a \bar\pi_b
-\frac12 R_{a\bar b c\bar d}\eta^{a\alpha}\bar\eta^b_\alpha\eta^{c\beta}\bar\eta^d_\beta +
Bg_{a\bar b}\eta^{a\alpha}{\bar\eta}^b_\alpha
\;:\quad \{Q^\alpha, \widetilde{\mathcal{H}}_0\}=0.
\ee
The last term in the Hamiltonians \eqref{hosup}, \eqref{hosup1} is obviously Zeeman term describing interaction of spin with an external magnetic field.
From the mathematical point of view, the shift in \eqref{hosup1} is the new $R$-symmetry $U(1)$ generator $\mathcal{R} =:
\frac12 g_{a\bar b}\eta^{a\alpha}\bar\eta^b_{\alpha}$. It extends $SU(2)$ $R$-symmetry generated by $\mathcal{R}^\alpha_\beta$ to $U(2)$ $R$-symmetry.
Since $\tilde{H}_0$ commutes with all other generators of the extended superalgebra, it can be interpreted as the central charge generator promoting the standard $su(2|1)$ superalgebra
to its central extension $\widehat{su}(2|1)$ \cite{ISKahler}.

All the generators  of $su(2|1)$ superalgebra (and of its central extension) are manifestly  invariant under the action of the isometry current \eqref{sNeth}:
\be
\{Q^\alpha, \mathcal{J}_\mu \}=\{{\overline Q}_\alpha, \mathcal{J}_\mu\}=\{\mathcal{R}^\alpha_\beta,\mathcal{J}_\mu\}=\{\mathcal{H}_{0},\mathcal{J}_\mu\}=0.
\ee
This means that the supersymmetric system constructed inherits all  the kinematical symmetries of the initial system.
In particular, in the case of $\mathbb{CP}^N$-Landau problem the extended system respects $SU(N+1)$ symmetry.

Thus we have accomplished the well defined ``weak  $\mathcal{N}=4$ supersymmetrization'' of the  Landau problem on a generic K\"ahler manifold
and found that its supersymmetry algebra is $\widehat{su}(2|1)$.

Finally, it is straightforward to write down the Lagrangian corresponding to \eqref{hosup},
\be
\mathcal{L}_{0}= g_{a\bar b}\dot z^a\dot{\bar z}^b +
i\frac{B}{2}({\partial_a K} \dot z^a-{\partial_{\bar a} K} \dot{\bar z}^a)+\frac{i}{2} g_{a\bar b}(\eta^{a\alpha} D_t\bar \eta^b_\alpha+\bar\eta^b_\alpha D_t \eta^{a\alpha})
+\frac12 R_{a\bar b c\bar d}\eta^{a\alpha}\bar\eta^b_\alpha\eta^{c\beta}\bar\eta^d_\beta - \frac B2
 g_{a\bar b}\eta^{a\alpha}{\bar\eta}^b_\alpha\,.
\label{lagr4landau}\ee
The Lagrangian corresponding to the shifted Hamiltonian \eqref{hosup1} is obviously $\widetilde{\mathcal{L}}_{0}=
\mathcal{L}_{0}-\frac{B}{2} g_{a\bar b}\eta^{a\alpha}\bar\eta^b_{\alpha}$. These Lagrangians provide a higher-dimensional  generalization of those constructed
in \cite{ivanov}, \cite{ivanovsidorov}, using the ${SU}(2|1)$ superfield techniques. The superfield derivation of \eqref{lagr4landau} will be given in Sect. VI.
The relevant ${SU}(2|1)$ off-shell multiplet content is $N$ chiral multiplets $({\bf 2, 4, 2})$. Note that the Lagrangian and Hamiltonian $\mathcal{L}_{0}$ and $\mathcal{H}_0$
coincide with the previously derived general expressions \eqref{lagrgen} and \eqref{GenH} for $\alpha = 1, 2$ and the choice $\mathcal{U} =
\frac12 R_{a\bar b c\bar d}\eta^{a\alpha}\bar\eta^b_\alpha\eta^{c\beta}\bar\eta^d_\beta -
Bg_{a\bar b}\eta^{a\alpha}{\bar\eta}^b_\alpha\,$.

\subsection{ $SU(4|1)$ (deformed $\mathcal{N}=8$)  supersymmetric Landau problem}
In the previous subsection  we  considered the coupling of $\mathcal{N}=4$ supersymmetric particle on K\"ahler manifold to a constant magnetic field
 and showed that the resulting system  yields the deformed $SU(2|1)$ supersymmetric Landau problem and that the latter inherits the whole isometry group
 of the original system. Now we perform a similar construction for ${\cal N}=8$ supersymmetric mechanics on  the {\sl special K\"ahler manifolds  of the rigid type} \cite{n8}.

The special K\"ahler  manifold of the rigid type is the K\"ahler manifold equipped  with the
symmetric tensor $f_{abc}dz^adz^bdz^c$ and its complex conjugate
which obey the following  compatibility conditions:
\be \frac{\partial}{\partial{\bar z}^d}f_{abc}=0\; ,\qquad f_{abc;d}=f_{abd;c}\; ,\qquad
\qquad R_{a\bar b c\bar d}=-{\bar f}_{\bar b \bar d {\bar n}}
g^{{\bar n}m }f_{mac}\,, \label{comp}\ee
where  $f_{abc;d}=f_{abc,d}-
\Gamma^e_{da}f_{ebc}-\Gamma^e_{db}f_{aec}-\Gamma^e_{dc}f_{abe}
$ is  the
covariant derivative of the third-rank  covariant  tensor. The special K\"ahler manifolds of the rigid type are widely known
 because of their close relevance to T-duality that relates the UV and IR limits of ${\cal N}=2, d=4$ super Yang-Mills theory \cite{SW}.

To construct the relevant supersymmetric Landau problem we choose the symplectic structure \eqref{ss} and Poisson brackets \eqref{sPB} with $su(4)$ spinor indices $\alpha,\beta=1,\ldots,4$.
To avoid possible confusion,  we relabel them by the capital Latin letters $I,J,K,L\,$.
With this notation, the ``flat'' $\mathcal{N}=8$ supersymmetry algebra reads
\be
\{Q^{I},Q^{J}\}=\{\bQ_I,\bQ_{J}\}= 0,\quad
\{Q^{I},\bQ_{J}\}= i\delta^I_J{\cal H}_{SUSY}. \label{188}
\ee
Following \cite{n8} we define  the supercharges as
\be
Q^{I}=\pi_a\eta^{aI}+\frac{i}{3}\,{\bar f}_{abc}
\,{\bar T}^{abcI},\quad \bQ_{I}={\bar\pi}_a{\bar\eta}^a_{I}+
\frac{i}{3}\,f_{abc}\,T^{abc}_{I},\qquad T^{abc}_{I}\equiv
\frac{1}{2}\,\varepsilon_{IJKL}\,\eta^{aJ}\eta^{bK}\eta^{cL}\,,\label{N8Q}
\ee
 where  the
symmetric tensor $f_{abc}$
obeys the relations
 \eqref{comp}
 \footnote{Here we introduced  the  antisymmetric symbol $\varepsilon^{IJKL}$ satisfying the following identities:
$$
\varepsilon^{1234}=\varepsilon_{1234}=1\,,\quad\varepsilon^{IJKL}\varepsilon_{IJKL}=24\,,\quad
    \varepsilon^{IJKL}\varepsilon_{IJKM}= 6\,\delta^L_M\,,\quad
    \varepsilon^{IJKL}\varepsilon_{IJMN}= 2\left(\delta^K_M\,\delta^L_N - \delta^K_N\,\delta^L_M\right),
$$
$$
\varepsilon^{IJKL}\varepsilon_{IMNP}= \delta^J_M\,\delta^K_N\,\delta^L_P - \delta^J_M\,\delta^K_P\,\delta^L_N
+ \delta^J_N\,\delta^K_P\,\delta^L_M - \delta^J_N\,\delta^K_M\,\delta^L_P + \delta^J_P\,\delta^K_M\,\delta^L_N - \delta^J_P\,\delta^K_N\,\delta^L_M\,.
$$
The highest-degree monomial of the Grassmann  variables  can be represented as $
    \psi^I\psi^J\psi^K\psi^L = \frac{1}{24}\,\varepsilon^{IJKL}\,\big(\varepsilon_{MNPR}\,\psi^M\psi^N\psi^P\psi^R$\big).}.  Also,  we introduce the following
    deformation of the Poisson brackets used in \cite{n8}:
\be
\{\pi_a, z^b\}=\delta^b_a,\quad
\{\pi_a,\eta^{bI}\}=-\Gamma^b_{ac}\eta^{cI},\quad
\{\pi_a,\bar\pi_b\}=i(Bg_{a\bar b}- R_{a\bar b c\bar d}\eta^{cI}{\bar\eta}^d_I),
\quad
\{\eta^{aI}, \bar\eta^b_J\}=
i g^{a\bar b}\delta^{I}_{J}.
\ee
Then we can construct $R$-symmetry charges forming $su(4)$ algebra by the same relations as in the undeformed case,
\be
R^{I}_{J}=\eta^{aI}g_{a\bar b}\,\bar\eta^b_{J}- \frac{\delta^{I}_J}{4}\,\eta^{aK}g_{a\bar b}\,\bar\eta^b_{K},\qquad \{R^{I}_{J}, R^{K}_{L}\}=i\left(\delta^{K}_{J}R^{I}_{L}-\delta^{I}_{L}R^{K}_{J}\right).\label{N8R}
\ee
Calculating the modified Poisson brackets between the supercharges and $R$-charges, we arrive at the generators $\mathcal{H}_{SUSY}, Q^{I}, R^{I}_{J}$
which form the superalgebra  $su(4|1)$
\be
\begin{array}{c}
\{Q^{I},Q^{J}\}=\{\bQ_I,\bQ_{J}\}= 0,\quad
\{Q^{I},\bQ_{J}\}= i \delta^I_J{\cal H}_{0}+i B R^I_J,\\[4mm]
\{R^{I}_{J},Q^{K}\}=i \delta^K_J Q^{I} - \frac{i }{4}\,\delta^I_J Q^{K},\quad \{{\cal H}_{0},Q^{K}\}=-\frac{3i B}{4}\,Q^{K}.
\end{array}
\ee
Here,
\be\label{N8H}
{\cal H}_{0}=
g^{\bar a b}\bar\pi_a\pi_b + R_{a\bar bc\bar d} \Lambda_0^{a c\bar b\bar d}+\frac{B}{4}\,\eta^{aK}g_{a\bar b}\bar\eta^b_{K}- \frac 13 f_{abc;d}\Lambda^{abc d} -\frac
13{\bar f}_{abc;d}\bar\Lambda^{abcd}\,,\ee
where, as before, $f_{abc;d}$ is  the
covariant derivative of the third-rank  covariant symmetric tensor, and
\be
\Lambda^{abcd}:= -\frac 18\,
\varepsilon_{IJKL}\,\eta^{aI}\eta^{bJ}\eta^{cK}\eta^{dL},\quad
\Lambda_0^{ac\bar b\bar d}:=\frac{1}{2}\,\eta^{aI}\eta^{cJ}\bar{\eta}^{b}_I
\bar{\eta}^{d}_J.\ee
We observe that the inclusion of constant magnetic field $B$ deforms $\mathcal{N}=8, d=1$ Poincar\'e superalgebra to the $su(4|1)$ superalgebra.

Let us require that the isometry of K\"ahler structure given by the vector field ${\bf V}_\mu $   preserves as well   the third-order tensor $f_{abc}dz^adz^bdz^c$, {\sl i.e.} that
 the Lie derivative of the latter  along this field equals to zero:
\be
\mathcal{L}_{{\bf V}_\mu}f_{abc}dz^adz^bdz^c=0\quad \Leftrightarrow\quad  3V^d_{\mu,(b}f_{ac)d}+V^d_\mu f_{abc,d}=0.\label{fabc}
\ee
Using these additional relations,
one  can check that  the isometry generator  \eqref{sNeth} commutes with all elements of $SU(4|1)$ superalgebra:
\be\{\mathcal{J}_\mu, Q_I\}=\{\mathcal{J}_\mu, \bQ_I\}=\{\mathcal{J}_\mu, R^I_J\}=\{\mathcal{J}_\mu, \mathcal{H}_{\rm Lan}\}=0\,.
\ee
Thus we managed to define the consistent $SU(4|1)$ Landau problem on special K\"ahler manifolds of the rigid type.

In contrast to $SU(2|1)$ Landau problem we cannot bring the Hamiltonian to the form in which it commutes with the supercharges, except
for the trivial case $f_{abc} = 0$\,.

Finally, taking into account the correspondence \eqref{lagrgen}, we can write the expression for the relevant Lagrangian
\bea
{\cal L}_{0}&=&g_{a\bar b}{\dot z}^a{\dot {\bar z}}^b+
i\frac{B}{2}({\partial_a K} \dot z^a-{\partial_{\bar a} K} \dot{\bar z}^a)+
+\frac{i}{2} g_{a\bar b}(\eta^{aI}{D_t\bar\eta^{b}_{I}}+
 \bar\eta^b_{I}{D_t\eta^{aI}})-\frac{B}{4}\,\eta^{aK}g_{a\bar b}\bar\eta^b_{K}
\nonumber\\
&&+\,\frac13( f_{abc;d}\Lambda^{abcd} +{\bar f}_{\bar a\bar b\bar c;\bar d}\bar\Lambda^{\bar a\bar b\bar c\bar d}) +
f_{abc}g^{c\bar c'}{\bar f}_{\bar c' \bar d\bar e}\Lambda_0^{ab\bar d\bar e}.
\label{lag8Lan}
\eea

{The  re-derivation of this Lagrangian from the appropriate off-shell $SU(4|1)$ superfield formalism is given in Sect. \ref{SFsection}, where the conditions \eqref{comp} are resolved,
 in the {\sl special coordinate frame},  through the single holomorphic function $\mathcal{F}(z)$ known as Seiberg-Witten prepotential:
\be
g_{a\bar b}= \frac{\partial^2\mathcal{F}(z)}{\partial z^a\partial z^b} +c.c.,\qquad \Gamma_{ab{\bar c}}=\frac{\partial^3\mathcal{F}}{\partial z^a\partial z^b\partial z^c}\qquad  f_{abc}={\rm e}^{i\nu}\frac{\partial^3\mathcal{F}(z)}{\partial z^a\partial z^b\partial z^c}.
\label{local}
\ee
Clearly, the function $\mathcal{F}(z)$ is defined up to redefinition
\be
\mathcal{F}(z)\quad\to\quad \mathcal{F}(z)+ ic_{ab}z^a z^b+ c_a z^a+c,
\label{Fu}\ee
where $c_a, c$ being arbitrary complex constants, and $c_{ab}$ are real ones, $\overline{c}_{ab}=c_{ab}$.

The corresponding  K\"ahler potential is given by the expression
\bea
    K\left(z,\bar{z}\right)=\bar{z}^a\,\frac{\partial{\cal F}\left(z\right)}{\partial z^a}
    +z^a\,\frac{\partial\bar{\cal F}\left(\bar{z}\right)}{\partial \bar{z}^a}\,.
    \label{localK}
\eea
In these coordinates, the T-duality transformation is realized as follows \cite{SW}
\be
\left( z^a,\; \mathcal{F}(z)\right)\to
\left( u_a=\frac{\partial\mathcal{F}}{\partial z^a},\;\widetilde{\mathcal{F}}(u)
\right),
\quad {\rm where}\quad
 \frac{\partial^2\widetilde{\mathcal{F}}(u)}{\partial u_a\partial u_c}
\frac{\partial \mathcal{F}}{\partial z^c\partial z^b}=-\delta^a_b\;, \quad  \widetilde{\mathcal{F}}(u)=(u_az^a-\mathcal{F}(z))\vert_{u_a=\partial_a\mathcal{F} (z)}.
\label{duality}\ee}

\section{ $SU(2|1)$ K\"ahler superoscillator}
The K\"ahler oscillator  is defined  by the symplectic structure \eqref{ssB}  and  the Hamiltonian \cite{Kahlerosc}
\be\label{Kahosc1}
{H}_{osc}=g^{\bar a b}\left({\bar \pi}_a\pi_b +|\omega|^2\partial_{\bar{a}}K\partial_b K \;\right),
\ee
where $K(z,\bar z)$ is the K\"ahler potential.

This system is distinguished in that it is ``friendly'' to supersymmetrization: the addition of the potential \eqref{Kahosc1} amounts to minor changes
in the procedure of $SU(2|1)$ supersymmetrization of the Landau problem described in the previous section.
Namely, we can preserve the form \eqref{TR} of  $SU(2)$  $R$-charges   and adopt the following slightly modified expressions for the supercharges
\footnote{We use here the following rules for complex conjugation and raising and lowering of $SU(2)$ spinor indices
$$
\overline{\varepsilon_{\alpha \beta}}=-\varepsilon^{\alpha \beta},\qquad \varepsilon^{\alpha\beta}=-\varepsilon^{\beta \alpha},
\qquad \varepsilon_{12} = \varepsilon^{21}=1 ,\qquad \varepsilon^{\alpha\beta}\varepsilon_{\gamma\delta}=\delta^{\alpha}_{\delta}\delta^{\beta}_{\gamma}
-\delta^{\alpha}_{\gamma}\delta^{\beta}_{\delta}\,.
$$}
\be
\Theta^\alpha=\pi_a \eta^{a\alpha}+ i\bar\omega \bar\partial_a K  \varepsilon^{\alpha \beta}{\bar \eta}^{a}_{\beta},\quad \overline{\Theta}_{\alpha}=
\bar\pi_a \bar\eta^a_\alpha + i\omega\partial_a K\varepsilon_{\alpha \beta} {\eta}^{a \beta} .
\ee
Calculating their Poisson brackets, we obtain
\be
 \{\Theta^\alpha,\overline\Theta_\beta\}=i \delta^{\alpha}_{\beta}\mathcal{H}_{osc}+ i B \mathcal{R}^{\alpha}_{\beta},\quad \{\Theta^\alpha,\Theta^\beta\}=2i\bar \omega\mathcal{R}^{\alpha\beta},\quad \{\Theta^\alpha, \mathcal{R}^\beta_\gamma\}= - i\delta^{\alpha}_{\gamma}  \Theta^\beta+\frac{i}{2}\delta_{\gamma}^{\beta}  \Theta^{\alpha} ,
\label{oscalgsu21}\ee
where the Hamiltonian is now given by the expression
\be
{\cal H}_{osc}=g^{\bar  a b}(\bar\pi_a\pi_b +|\omega|^2\partial_{\bar a}K\partial_b K)
-\frac 12  R_{a\bar b c\bar d}\eta^{a\alpha}\bar\eta^b_\alpha\eta^{c\beta}\bar\eta^d_\beta
-\frac{1}{2}\omega K_{a;b}\eta^{a\alpha}\eta^b_\alpha -
\frac{1}{2}\bar \omega K_{\bar a;\bar b}\bar\eta^{a}_{\alpha}\bar\eta^{b\alpha} + \frac B2
 g_{a\bar b}\eta^{a\alpha}{\bar\eta}^b_\alpha\,. \label{hosuposc}
\ee
To close the superalgebra, we have to complete \eqref{oscalgsu21} by  the $SU(2)$ algebra relations between $R$-charges as is given in  \eqref{TR}, and by  the
 full set of Poisson brackets involving   the supercharges $\overline\Theta^\beta$.

In order to bring this superalgebra into the conventional form   it is convenient to  rotate the supercharges as
\be
Q^{\alpha}=e^{i\nu/2}\cos\lambda \Theta^{\alpha}+e^{-i\nu/2}\sin\lambda \varepsilon^{\alpha \gamma}\overline
\Theta_{\gamma} ,\qquad \overline Q_{\alpha}=e^{-i\nu/2}\cos\lambda \overline\Theta_{\alpha}-e^{i\nu/2}\sin\lambda \varepsilon_{\alpha \gamma}\Theta^{\gamma}\,,
\ee
where
\be
 \cos 2\lambda=\frac{B}{\sqrt{4|\omega|^2+B^2}},\quad
\sin 2\lambda=-\frac{2|\omega|}{\sqrt{4|\omega|^2+B^2}},
\qquad
\omega=|\omega|e^{i\nu}\,.
\label{2lambda}\ee
In terms of these newly defined quantities  the symmetry algebra is rewritten as
\bea
&\{Q^{\alpha},\overline Q_{\beta}\}=i \delta^{\alpha}_{\beta}{\cal H}_{osc}+\sqrt{4|\omega|^2+B^2} \  \mathcal{R}^{\alpha}_{\beta},\quad
\{Q^\alpha,\mathcal{H}_{osc}\}=\frac{i}{2} \sqrt{4|\omega|^2+B^2}\ Q^\alpha, \quad \{Q^{\alpha},Q^{\beta}\}=\{\overline Q_{\alpha},\overline Q_{\beta}\}=0,&\\[4mm]
&\{Q^{\alpha},\mathcal{R}^{\beta}_{\gamma}\}= - i\delta^{\alpha}_{\gamma}  Q^\beta+\frac{i}{2}\delta_{\gamma}^{\beta}  Q^{\alpha}
 \quad   \{\mathcal{R}^\alpha_\beta,\mathcal{R}^\gamma_\delta\}=i\delta^\gamma_\beta \mathcal{R}^\alpha_\delta-i \delta^\alpha_\delta\mathcal{R}^\gamma_\beta
\quad
\{\mathcal{R}^\alpha_\beta,\mathcal{H}_{osc}\}=0.&
\eea
Comparing these relations with those of the supersymmetric $\mathcal{N}=4$ Landau problem \eqref{QR}, we can identify them as defining
$SU(2|1)$ superalgebra with the deformation parameter $m=\sqrt{4|\omega|^2+B^2}$\,.

The Lagrangian of  $SU(2|1)$ supersymmetric K\"ahler oscillator is given by the general expression \eqref{lagrgen}, with
\be
\mathcal{U}=
|\omega|^2 g^{a\bar b}\partial_a K \partial_{\bar b} K
-\frac 12  R_{a\bar b c\bar d}\eta^{a\alpha}\bar\eta^b_\alpha\eta^{c\beta}\bar\eta^d_\beta
-\frac{\omega}{2} K_{a;b}\eta^{a\alpha}\eta^b_\alpha -
\frac{\bar \omega }{2}K_{\bar a;\bar b}\bar\eta^{a}_\alpha\bar\eta^{b\alpha} + \frac B2
g_{a\bar b}\eta^{a\alpha}{\bar\eta}^b_\alpha .
\label{supot}\ee
The  supersymmetrization procedure described above  is capable to produce  a   family of non-equivalent Hamiltonians parameterized by an arbitrary holomorphic function.
Namely, replacing the initial K\"ahler potential $K$  by the gauge-equivalent one,
\be
K(z,\bar z)\to K(z,\bar z)+\frac{1}{\omega}U(z)+\frac{1}{\bar \omega}{\bar U}(\bar z),
\label{gauge}\ee
we obtain the class of Hamiltonians  parameterized by an arbitrary holomorphic function $U(z)$,
\bea
\mathcal{H}_{osc}\to{\cal H}_{osc}&=&g^{\bar  a b}(\bar\pi_a\pi_b + \partial_{\bar a} \bar{U} \partial_b U     ) - \frac12 R_{a\bar b c\bar d}\eta^{a\alpha}\bar\eta^b_\alpha \eta^{c\beta}\bar\eta^d_\beta +\frac{1}{2}U_{a;b}\eta^{a\alpha}\eta^b_\alpha+\frac{1}{2} {\bar U}_{\bar a;\bar b}\bar\eta^a_\alpha\bar\eta^{b\alpha}+ \frac B2
g_{a\bar b}\eta^{a\alpha}{\bar\eta}^b_\alpha\nonumber\\
&&+|\omega|^2 g^{\bar a b}\partial_{\bar a}K\partial_b K
+|\omega| g^{\bar a b}\left(\partial_{\bar a} K \partial_b {U} +\partial_{\bar a} \bar{U} \partial_b K\right)-\frac{\omega}{2} K_{a;b}\eta^{a\alpha}\eta^b_\alpha -
\frac{\bar \omega }{2}K_{\bar a;\bar b}\bar\eta^a_\alpha\bar\eta^{b\alpha} . \label{hosupkah}
\eea
In the  limit $\omega=0$ we  arrive at  the well-known Hamiltonian which admits, in the absence of magnetic field,
the ``flat'' $\mathcal{N}=4$ supersymmetry (see, e.g. \cite{PRDrapid}). It is given by the first line in the above expression with $B=0$\\


\begin{center}
{\bf  A. Two-dimensional anisotropic oscillator}
\end{center}
The supersymmetrization procedure outlined above makes it possible to extend the class
of the known systems admitting such a supersymmetrization. Here we illustrate this on the case of  two-dimensional harmonic oscillator
which is the simplest system possessing the conventional $\mathcal{N}=4, d=1$ ``Poincar\'e'' supersymmetric extension. Take the one-dimensional
complex space $(\mathbb{C}, \;ds^2 = dz\, d\bar z)$ and consider on it the K\"ahler oscillator defined by the potential
\be
K(z,\bar z)=z\bar z+\frac{i gz^2}{2\omega}-\frac{i \bar g{\bar z}^2 }{2\bar \omega}.
\ee
It gives rise to the  following K\"ahler-oscillator system
\be
{H}=\pi\bar\pi+ (\omega\bar\omega+ g\bar g)z\bar z+i\bar\omega gz^2 - i\omega \bar g{\bar z}^2, \quad \{\pi, z\}=\{\bar\pi, z\}=1,\quad \{\pi,\bar\pi\}=i B.
\ee
Diagonalizing this potential,  we arrive at the  two-dimensional anisotropic oscillator system with frequencies
 \be
\omega^\pm=\Big| |\omega|\pm |g| \Big |\,.
\ee
For  the choice $\omega=0$ it yields the two-dimensional isotropic oscillator with the frequency $|g|$,
 which admits, in the absence of magnetic field, the standard $\mathcal{N}=4, d=1$ supersymmetrization. In the presence of magnetic field this
 supersymmetry is deformed to $SU(2|1)$. In the opposite limit, at $g=0$, we once again obtain some $SU(2|1)$ supersymmetric extension of two-dimensional isotropic oscillator,
 but different from the first option.
In the  generic case of $g\neq 0, \omega\neq 0$  the  procedure proposed allows to construct $SU(2|1)$ superextension of the
two-dimensional {\bf  anisotropic} oscillator interacting with a constant magnetic field perpendicular to the plane.
Enlarging the above set of Poisson brackets by the relation $\{\eta^\alpha,\bar\eta_\beta\}=i\delta^\alpha_\beta$, we
can write down the Hamiltonian of the supersymmetric extension of this system as
\be
\mathcal{H}_{anosc}=\pi\bar\pi+ (\omega\bar\omega+ g\bar g)z\bar z+i\bar\omega gz^2-i\omega \bar g\bar{z}^2
-\frac{ig}{2} \eta^{\alpha}\eta_{\alpha}+\frac{i\bar{g}}{2}\bar \eta_{\alpha}\bar \eta^{\alpha}+\frac B2 \eta^{\alpha}\bar\eta_\alpha.
\ee
The relevant  supercharges and $R$-charges have the following simple form
\be
\Theta^\alpha=\pi \eta^{\alpha}+ (i\bar \omega z + \bar g \bar z) {\varepsilon^{\alpha \beta}}{\bar \eta}_{\beta} \qquad
\mathcal{R}^\alpha_\beta= \eta^{\alpha}\bar\eta_\beta - \frac{1}{2}\delta^\alpha_\beta \eta^{\gamma}\bar\eta_\gamma.
\ee
It is straightforward to extend this model  to $N$-dimensional complex Euclidian space $\mathbb{C}^N\,$ (see Section VI).
\section{Superfield formulation}\label{SFsection}
The one-particle (i.e. one-(complex)dimensional)  versions of the Lagrangians presented above were derived from the $SU(2|1)$ and $SU(4|1)$ superfield approaches
in \cite{ISKahler} and \cite{ILS19rev}.
The generalization of these models to the $N$-dimensional
case is straightforward. We briefly describe it below.
\subsection{$SU(2|1)$ case}

As the first step, we reproduce the  Lagrangian of $SU(2|1)$ K\"ahler  superoscillator corresponding to \eqref{hosuposc}, and its particular case, the Lagrangian
of $SU(2|1)$ supersymmetric Landau problem \eqref{lagr4landau}.

In \cite{ivanovsidorov} and \cite{ISKahler} the coset method was used to define the world-line realizations of the supergroup $SU(2|1)$ on the $d=1$ superspace
$(t, \theta_\alpha, \bar\theta^\beta)$ identified with the coset of $SU(2|1)$ over its $R$-symmetry subgroup $SU(2)$. The basic objects of this realization
are covariant spinor derivatives
\bea
    &&{\cal D}^\alpha=e^{-\frac{i mt}{2}}\left[\left(1+\frac{m}{2}\,\bar{\theta}^\beta\theta_\beta
    -\frac{3m^2}{16}\left(\bar{\theta}^\beta\theta_\beta\right)^2\right)\frac{\partial}{\partial\theta_\alpha}
    - \frac{m}{2}\,\bar{\theta}^\alpha\theta_\beta\frac{\partial}{\partial\theta_\beta}-\frac{i}{2}\,\bar{\theta}^\alpha\partial_t\right],\nn
    &&\bar{{\cal D}}_\alpha =e^{\frac{i mt}{2}}\left[-\left(1+\frac{m}{2}\,\bar{\theta}^\beta\theta_\beta
    -\frac{3m^2}{16}\left(\bar{\theta}^\beta\theta_\beta\right)^2\right)\frac{\partial}{\partial\bar{\theta}^\alpha}
    + \frac{m}{2}\,\bar{\theta}^\beta\theta_\alpha\frac{\partial}{\partial\bar{\theta}^\beta}+\frac{i}{2}\,\theta_\alpha\partial_t\right],
\eea
which, in the contraction limit $m=0$, become standard covariant spinor derivatives of flat ${\cal N}=4, d=1$ supersymmetry. The chiral $SU(2|1)$
superfields  $\Phi^a(t,\hat{\theta},\bar\hat{\theta})$ satisfy the generalized $SU(2|1)$ covariant chirality constraints \cite{ISKahler}
\bea
    \left(\cos{\lambda}\,\bar{{\cal D}}_\alpha -\sin{\lambda}\,{\cal D}_\alpha\right)\Phi^a = 0\,.\label{GenChiral}
\eea
In the appropriate superspace basis the conditions \eqref{GenChiral} become ``short'' up to an overall factor,
\bea
    \left(\cos{\lambda}\,\bar{{\cal D}}_\alpha -\sin{\lambda}\,{\cal D}_\alpha\right)\Phi^a =
    \left[1+\frac{B}{4}\,\bar{\hat{\theta}}^\beta\hat{\theta}_\beta+\frac{\omega}{4}\,\left(\hat{\theta}_\beta\hat{\theta}^\beta
    +\bar{\hat{\theta}}^\beta\bar{\hat{\theta}}_\beta\right)
    -\frac{m^2}{32}\left(\bar{\hat{\theta}}^\beta\hat{\theta}_\beta\right)^2\right]
    \left[-\frac{\partial}{\partial\bar{\hat{\theta}}^\alpha}+\frac{i}{2}\,\hat{\theta}_\alpha\partial_t\right]\Phi^a\,,
\eea
and are solved by the expressions
\bea
    \Phi^a (t_{\rm L},\hat{\theta}_\alpha)=z^a+\hat{\theta}_\alpha\eta^{a\alpha} +\frac{1}{2}\,\hat{\theta}_\alpha\hat{\theta}^\alpha A^a ,\qquad t_{\rm L} = t +\frac{i}{2}\, \bar{\hat{\theta}}^\alpha\hat{\theta}_\alpha \,.
\eea
The dependence on the new parameter $\lambda$ is now hidden in the definition of the superspace coordinates $t_{\rm L}$ and $\hat{\theta}_\alpha\,$,
which have the following $SU(2|1)$ transformation properties
\bea
    &&\delta\hat{\theta}_{\alpha}=\cos{\lambda}\left(\epsilon_{\alpha}\,e^{\frac{i}{2}m t_{\rm L}}
    +\frac{m}{2}\,\bar{\epsilon}^\beta\hat{\theta}_\beta\hat{\theta}_{\alpha}\, e^{-\frac{i}{2}m t_{\rm L}}\right)
    +\sin{\lambda}\left(\bar{\epsilon}_\alpha\, e^{-\frac{i}{2}m t_{\rm L}} +\frac{m}{2}\,\epsilon^\beta\hat{\theta}_\beta\hat{\theta}_{\alpha}\, e^{\frac{i}{2}m t_{\rm L}}\right),\\
    && \delta t_{\rm L} = i\cos{\lambda}\,\bar{\epsilon}^\beta \hat{\theta}_\beta\, e^{-\frac{i}{2}m t_{\rm L}} - i\sin{\lambda}\,
    \epsilon^\beta\hat{\theta}_\beta \,e^{\frac{i}{2}m t_{\rm L}}.
\eea
These coordinate transformations induce the off-shell $SU(2|1)$  supersymmetry transformation of chiral superfields. On the component fields they are realized as
\be
\begin{array}{c}
    \delta z^a = -\left(\cos{\lambda}\,\epsilon_\alpha\,e^{\frac{i}{2}m t} + \sin{\lambda}\,\bar{\epsilon}_\alpha\,e^{-\frac{i}{2}m t}\right)\eta^{a\alpha},\\
    \delta \eta^{a\alpha} =  \bar{\epsilon}^\alpha\left(i\cos{\lambda}\,\dot{z}^a-\sin{\lambda}\,A^a\right)
    e^{-\frac{i}{2}m t}-\epsilon^\alpha\left(i\sin{\lambda}\,\dot{z}^a + \cos{\lambda}\,A^a\right)e^{\frac{i}{2}m t},\\
    \delta A^a = -\,\cos{\lambda}\,\bar{\epsilon}_\alpha\left(i\dot{\eta}^{a\alpha}+\frac{m}{2}\,\eta^{a\alpha}\right)e^{-\frac{i}{2}m t}
    +\sin{\lambda}\,\epsilon_\alpha\left(i\dot{\eta}^{a\alpha}-\frac{m}{2}\,\eta^{a\alpha}\right)e^{\frac{i}{2}m t},
\end{array}\label{offsh}
\ee
where $\epsilon_\alpha$ are ``infinitesimal" Grassmann parameters.

The corresponding off-shell superfield Lagrangian  is as follows (see \cite{ISKahler} for one-particle case)
\bea
    {\cal L} = \int d^2\hat{\theta}\,d^2\bar{\hat{\theta}} \left[1+\frac{B}{2}\,\bar{\hat{\theta}}^\alpha\hat{\theta}_\alpha+\frac{\omega}{2}\,\left(\hat{\theta}_\alpha\hat{\theta}^\alpha
    +\bar{\hat{\theta}}^\alpha\bar{\hat{\theta}}_\alpha\right)\right]K\left(\Phi^a,\bar{\Phi}^b\right),\label{lagrSF}
\eea
where \footnote{We limit our attention to real frequencies $\omega = |\omega|$ in order to match the superfield approach elaborated in \cite{ISKahler}.
In fact, one can easily generalize this consideration to $\omega\in\mathbb{C}$.}
\be
    B=m\cos 2\lambda\,,\quad \omega=-\frac{m}{2}\sin 2\lambda\,.
\ee

 It is straightforward to check that the transformation of the factor within the square brackets in \eqref{lagrSF} precisely cancels the transformation of the integration
measure $dt_L d^2\hat{\theta} d^2\bar{\hat{\theta}}$.
Integrating in \eqref{lagrSF} over $\hat{\theta}, \bar{\hat{\theta}}$ and eliminating the auxiliary fields $A^a$, we recover the on-shell Lagrangian \eqref{lagrgen}
with the expression \eqref{supot} for ${\cal U}$. In the particular case $\lambda=0$ ($\omega=0$), we arrive at the Lagrangian \eqref{lagr4landau} of Landau problem.
Holomorphic terms \eqref{gauge} can be naturally inserted in \eqref{lagrSF} with $\omega \neq 0$ through the shift
\bea
    K\left(\Phi^a,\bar{\Phi}^b\right)\to K\left(\Phi^a,\bar{\Phi}^b\right)+\frac{1}{\omega}\,U(\Phi^a)+\frac{1}{\omega}\,{\bar U}(\bar{\Phi}^b),\label{KUU}
\eea
which amounts to introduction of the additional superpotential terms  which, in components, induce the modified
potential ${\cal U}\,$,  as in \eqref{hosupkah}.

It is instructive to see how the phenomenon of preserving the isometries under the deformation manifests itself in the superfield language.
 For this purpose  we  need to know how the K\"ahler potential itself transforms under isometry of K\"ahler structure given by \eqref{iso}.
 To this end, we rewrite the equation $({\rm b})$ in \eqref{Killing} in the equivalent
form as
\be
\partial_c\partial_{\bar d} \big\{\big[V_\mu^{a}(z){\partial}_a+ {V}_\mu^{\bar a}(\bar z){\partial}_{\bar a}\big]K(z, \bar z)\big\} = 0\,, \label{TranK}
\ee
whence
\be
\big[V_\mu^{a}(z){\partial}_a+ {V}_\mu^{\bar a}(\bar z){\partial}_{\bar a}\big]K(z, \bar z) = \varphi_\mu(z) + \bar{\varphi}_\mu(\bar z)\,. \label{TranK2}
\ee
The holomorphic function $\varphi_\mu(z)$, in each specific case, can be defined up to a constant by differentiating \eqref{TranK2} with respect to $z^b$.

 The isometry transformations of
the K\"ahler manifold in the superfield coordinates are obtained just by the changes $z^a \rightarrow \Phi^a$, $\bar{z}^a \rightarrow \bar{\Phi}^a$ in the relevant holomorphic
Hamiltonian vector fields. Recalling the transformation \eqref{TranK2} of $K(z, \bar z)$ under isometry, we see that the superfield Lagrangian in \eqref{lagrSF} is transformed as
\be
\delta^*K = b^\mu\varphi(\Phi^a)_\mu  + \bar{b}^\mu\bar{\varphi}(\bar{\Phi}^a)_\mu\,, \label{PhibarPhi}
\ee
where $b_\mu, \bar{b}_\mu$ are constant isometry parameters. Taking the bar-spinor derivatives from the integration measure and making use of the chirality of $\Phi^a$,
it is easy to see that the holomorphic term in \eqref{PhibarPhi} does not contribute at $\omega = \lambda = 0, B=m$. The vanishing of the contribution from the
conjugated antiholomorphic term in \eqref{PhibarPhi} can be proved after passing to the right-chiral basis in the $SU(2|1)$ superspace. This is the superfield proof of
the property that the $SU(2|1)$ super Landau model inherits all isometries of the undeformed case $\omega = \lambda = m = 0$. The isometries
are not generically inherited by the K\"ahler superoscillator, when  $\omega \neq 0$.

It should be pointed out that the input parameters of the above superfield formalism are just the contraction mass-dimension parameter $m$ coming from
the (anti)commutation relations of the $su(2|1)$ algebra and the angle $\lambda$ coming from the chirality constraint \eqref{GenChiral}. The physical meaning of
these parameters as the strength of the external magnetic field and the oscillator frequency is revealed at the level of the component Lagrangians and Hamiltonians.
\subsection{$SU(4|1)$ case}

Next, let us present the $SU(4|1)$ superfield formulation for the Lagrangian of the ${\cal N}=8$ Landau problem \eqref{lag8Lan}, based on the superspace approach developed in \cite{SU41}.
This superfield Lagrangian is written in terms of chiral ${\bf(2,8,6)}$ superfields as follows (its one-particle case was constructed  in \cite{ILS19rev})
\be
    S= \int dt\,{\cal L} = -\int dt_{\rm L}\,d^4\theta\,e^{-3i m t_{\rm L}}\,\mathcal{F}\left(\Phi^a\right)
    - \int dt_{\rm R}\,d^4\bar{\theta}\,e^{3i mt_{\rm R}}\,\bar{\mathcal{F}}\left(\bar{\Phi}^{a}\,\right),\quad m=|B|.\label{N8SF}
\ee
Here $\mathcal{F}(z)$ is Seiberg-Witten prepotential \eqref{local}, while the $\theta$-expansion of the superfields $\Phi^a$ reads
\bea
  & \Phi^a\left(t_{\rm L},\theta_{I}\right) &= z^a + \theta_{K}\eta^{aK}e^{3i mt_{\rm L}/4} + \frac{1}{2}\,\theta_{I}\theta_{J}A^{aIJ}e^{3i mt_{\rm L}/2}
   -\frac{1}{6}\,\varepsilon^{IJKL}\theta_{I}\theta_{J}\theta_{K}\left(i\dot{\bar{\eta}}^a_L-\frac{m}{4}\,\bar{\eta}^a_L\right)e^{9i mt_{\rm L}/4}\nonumber\\
    &&+\,\frac{1}{24}\,\varepsilon^{IJKL}\,\theta_{I}\theta_{J}\theta_{K}\theta_{L}\left(\ddot{\bar{z}}^a+i m\dot{\bar{z}}^a\right) e^{3i mt_{\rm L}},
    \label{286} \eea
with the following conjugation rules
$\overline{\left(A^{aIJ}\right)}=A^a_{IJ}=\frac{1}{2}\,\varepsilon_{IJKL}\,A^{aKL},
 \overline{\left(\eta^{aI}\right)}=\bar{\eta}^a_I $.

The coordinate set $\{t_{\rm L}\,,\theta^I\}$ is closed under the $SU(4|1)$ transformations
\bea
    \delta\theta_{I}=\epsilon_{I}+m\,\bar{\epsilon}^{K}\theta_{K}\theta_{I}\,,\qquad \delta t_{\rm L}=i\bar{\epsilon}^K\theta_K\,.\label{left_tr}
\eea
The corresponding  off-shell supersymmetry transformations of the component fields read
\be
\begin{array}{c}
  \delta z^a = -\,\epsilon_{K}\eta^{aK}e^{3i mt/4},\qquad \delta \bar{z}^a = \bar{\epsilon}^{K}\bar{\eta}^a_{K} \,e^{-3i mt/4},\\
   \delta A^{aIJ} = 2\,\bar{\epsilon}^{[I}\left(i \dot{\eta}^{aJ]}+\frac{m}{4}\eta^{aJ]}\right)e^{-3i mt/4} + \varepsilon^{IJKL}\,\epsilon_{[K}\left(i \dot{\bar{\eta}}^a_{L]}-\frac{m}{4}\,\bar{\eta}^a_{L]}\right)e^{3i mt/4},\\
   \delta \eta^{aI} = \bar{\epsilon}^I\left(i\dot{z}^a\right)e^{-3i mt/4} - \epsilon_K\,A^{aIK}e^{3i mt/4},\quad\delta \bar{\eta}^{a}_I = -\,\epsilon_I\left(i \dot{\bar{z}}^a\right)e^{3i mt/4} - \bar{\epsilon}^K A^a_{IK}\,e^{-3i mt/4}.
    \end{array}\label{tr286}
\ee
Integration in \eqref{286} over $\theta$, $\bar{\theta}$ gives the off-shell Lagrangian
\bea
    {\cal L}_{\rm off-shell} &=& g_{a\bar{b}}\left[\dot{z}^a\dot{\bar{z}}^{b}-\frac{1}{4}\,A^{aIJ}A^b_{IJ}+ \frac{i}{2}\left(\eta^{aK}\dot{\bar{\eta}}^b_K-\dot{\eta}^{aK}\bar{\eta}^b_K\right)- \frac{m}{4}\,\eta^{aK}\bar{\eta}^b_K\right]-\frac{i}{2}\left(\dot{z}^c\,\partial_{c}g_{a\bar{b}} - \dot{\bar{z}}^c\,\partial_{\bar{c}}g_{a\bar{b}}\right)\eta^{aK}\bar{\eta}^b_K\nn
    &&+\,i m\left(\dot{z}^a\,\partial_{\bar a}\bar{\mathcal{F}}-\dot{\bar{z}}^a\,\partial_{a}\mathcal{F}\right)
    +\frac{1}{2}\,A^b_{IJ}\,\eta^{aI}\eta^{cJ}\,\partial_{c}g_{a\bar{b}}-\frac{1}{2}\,A^{aIJ}\bar{\eta}^{b}_I\bar{\eta}^{c}_J\,\partial_{\bar{c}}g_{a\bar{b}}\nn
    &&-\,\frac{1}{24}\left[\varepsilon_{IJKL}\,\eta^{aI}\eta^{bJ}\eta^{cK}\eta^{dL}\,\partial_{c}\partial_{d}g_{a\bar{b}}
    + \varepsilon^{IJKL}\,\bar{\eta}^a_I\bar{\eta}^b_J\bar{\eta}^c_K\bar{\eta}^d_L\,\partial_{\bar{c}}\partial_{\bar{d}}g_{a\bar{b}}\right],\label{L_286}
\eea
where the metric $g_{a\bar{b}}$ is identified with the metric defined in \eqref{local}.
The subsequent elimination of the auxiliary fields $A^{aIJ}$ yields just the on-shell Lagrangian \eqref{lag8Lan}.

{ It is important that the superfield action \eqref{N8SF} is invariant under the transformations corresponding to \eqref{Fu}(see \cite{dWVP})
\be
   \mathcal{F}\left(\Phi^a\right) \rightarrow \mathcal{F}\left(\Phi^a\right) + i c_{ab}\Phi^a\Phi^b + c_{a}\Phi^a+ c,\quad
  \bar{\mathcal{F}}\left(\bar{\Phi}^{a}\,\right) \rightarrow \bar{\mathcal{F}}\left(\bar{\Phi}^{a}\,\right) - i c_{ab}\bar{\Phi}^{a}\bar{\Phi}^{b} + \bar{c}_{a}\bar{\Phi}^{a} + \bar{c}\,,
\label{tr_F}\ee
where $c,c_a$ are complex numbers, and $c_{ab}$ are real ones.

These transformations are just the ${\cal N}=8$ superfield version of the general transformations of the holomorphic prepotential ${\cal F}(z)$
under an arbitrary {\sl isometry of the special K\"ahler  structure}, i.e. of the isometry of  K\"ahler structure preserving  holomorphic third-order tensor \eqref{fabc}
 (see Appendix A). Hence,   the invariance of \eqref{N8SF} under \eqref{tr_F} explicitly demonstrates
that the deformed ${\cal N}=8$ supersymmetric mechanics we are considering inherits the full set of isometries of the undeformed case.

The proof of this superfield invariance is not too easy. To this end, one needs to represent the invariant chiral measure $d^4\theta\,e^{-3i m t_{\rm L}}$
in the action \eqref{N8SF} in terms of covariant derivatives (up to total time derivatives) as\footnote{Though expressions for $SU(4|1)$ covariant derivatives were not calculated, the function ${\cal D}^I{\cal D}^J{\cal D}^K{\cal D}^L{\cal F}\left(\Phi^a\right)$ is $SU(4|1)$ invariant. Hence, it must give the same invariant action \eqref{N8SF}.}
\bea
    d^4\theta\,e^{-3i m t_{\rm L}} = \frac{1}{24}\,e^{-3i m t_{\rm L}}\,\varepsilon_{IJKL}\,\partial^I\partial^J\partial^K\partial^L = \frac{1}{24}\,\varepsilon_{IJKL}\,{\cal D}^I{\cal D}^J{\cal D}^K{\cal D}^L.\label{D4}
\eea
Covariant derivatives anticommute as
\be
    \left\lbrace\bar{\cal D}_{I},\bar{\cal D}_{J}\right\rbrace = 0\,,\qquad\left\lbrace{\cal D}^{I},{\cal D}^{J}\right\rbrace = 0\,,\qquad\left\lbrace{\cal D}^{I},\bar{\cal D}_{J}\right\rbrace= \delta^I_J{\cal H}_{0} + m \tilde{R}^I_J\,,\qquad
    \tilde{R}^I_J {\cal D}^{K} = \frac{1}{4}\,\delta^I_J {\cal D}^{K} - \delta^K_J{\cal D}^{I},
\ee
where $\tilde{R}^I_J$ are $SU(4)$ matrix generators acting on external indices of superfields and covariant derivatives.
The chiral superfield $\Phi^a$ ($a=1, \ldots N$) describing $N$ multiplets ${\bf (2,8,6)}$ satisfies the constraints \cite{dWVP}
\bea
    {\cal D}^I \bar{\Phi}^a = 0\,,\qquad \bar{\cal D}_K \Phi^a = 0\,,\qquad {\cal D}^I{\cal D}^J \Phi^a = \frac{1}{2}\,\varepsilon^{IJKL}\,\bar{\cal D}_K\bar{\cal D}_L \bar{\Phi}^a.\label{c_286}
\eea
Exploiting \eqref{D4}-\eqref{c_286} for the action \eqref{N8SF}, one can show its invariance under the transformations \eqref{tr_F}. Another, more direct proof is to substitute
the explicit expressions \eqref{286} for $\Phi^a$ and the conjugated ones for $\bar{\Phi}^a$ into \eqref{tr_F} and to be convinced that the coefficients of the higher-order
monomials in $\theta_I (\bar{\theta}^I)$ in the holomorphic(antiholomorphic) shifts \eqref{tr_F} either are combined into total $t$-derivatives or just vanish. Note that
the reality condition for the coefficient $c_{ab}$ in \eqref{tr_F} is essential for ensuring the properties just mentioned.

Derivation of the purely bosonic counterpart of the transformations \eqref{tr_F} from the isometry condition \eqref{fabc} is discussed in Appendix A.}

\section{Examples of superintegrable  K\"ahler oscillator models}
In the previous sections we dealt with  two classes of models admitting deformed supersymmetry: the Landau problems, and the  K\"ahler oscillators.
In the case of Landau problem we found that the supersymmetric extensions preserve all (kinematical) symmetries of the initial systems.
But we  were not able to prove the similar general proposition for the K\"ahler oscillators.
In this section we present supersymmetric extensions of two particular types of the K\"ahler oscillator systems which possess  kinematical symmetries and the hidden symmetries generated  by the constants of motion quadratic in momenta.
These two types are encompassed by the following models
\begin{itemize}
\item $\mathbb{C}^N$-oscillator (the sum of $N$ two-dimensional isotropic oscillators) and  $\mathbb{C}^N$-Smorodinsky-Winternitz system (the sum of $N$ copies
 of two-dimensional isotropic oscillators deformed by  ring-shaped potentials).
 \item $\mathbb{CP}^N$-oscillator and $\mathbb{CP}^N$-Rosochatius system, which are  superintegrable counterparts of $\mathbb{C}^N$-oscillator
 and $\mathbb{C}^N$-Smorodinsky-Winternitz systems on the complex projective spaces.
\end{itemize}
Our main goal will be to inspect whether $SU(2|1)$ supersymmetric extensions of these systems inherit their hidden symmetries.
\subsection{Euclidean spaces}
We start by considering the K\"ahler oscillators  on the complex Euclidian space  $(\mathbb{C}^N, ds^2 =\sum_{a=1}^N dz^a d{\bar z}^a)$.
The relevant phase space is defined by the Poisson brackets
\be
\{\pi_a, z^b\}=\delta^b_a,\quad \{\bar \pi_a, \bar z^b\}=\delta^b_a,\quad\{\pi_a, \bar\pi_b\}=i B \delta_{a\bar b}\,.
\label{canonical}\ee
The set of symmetries of this space is constituted by the $SU(N)$ generators
\be
J_{a\bar b}=i\pi_a z^b-i\bar{\pi}_b\bar{z}^a-{B}z^b{\bar z}^a\;:
\{J_{{\bar a} b}, J_{\bar c d}\}=
i\delta_{\bar a d}J_{\bar b c}
-i\delta_{\bar c b}J_{\bar a d},
\label{suN}\ee
and  the translation generators
\be
J_{a}=i \pi_a-B\bar z^a
\;:\quad\{J_{a},J_b\}=
\{J_{a},{\bar J}_b\}=0,\quad\{J_a, J_{b\bar c}\}= -i J_b\delta_{a\bar c}.
\ee

For the construction of $SU(2|1)$ supersymmetric K\"ahler oscillator models on this space  we have to complete the Poisson brackets \eqref{canonical} by the following ones
\be
\{\eta^{a\alpha},\bar\eta^b_{\beta}\}=i \delta^{a\bar b}\delta^\alpha_\beta,
\ee
with $\alpha,\beta=1,2$.
Then we should perform the  $SU(2|1)$ supersymmetrization procedure described above, for  the appropriate choice of the initial bosonic
K\"ahler oscillator model.
\subsubsection*{Harmonic oscillator}

We define  the $\mathbb{C}^N$-harmonic oscillator defined as a K\"ahler oscillator with $K(z,\bar z)=\sum_{a=1}^Nz^a\bar z^a$  and $\omega=\bar\omega$:

\be
{H}_{osc}=  \sum_{a=1}^N\Big(\pi_a\bar\pi_a+ \omega^2 z^a\bar z^a\Big).
\ee
This system possesses $SU(N)$ kinematical symmetry generated by the generators \eqref{suN}, and hidden symmetries defined by the so-called Fradkin tensor
\be
I_{a\bar b}=\pi_a\bar\pi_b+\omega^2\bar z^a z^b\; :\{I_{a\bar b},I_{c\bar d}\}=i \delta_{a\bar d}J_{c\bar b}-
i \delta_{c\bar b}J_{a\bar d},\quad \{I_{a\bar b},J_{c\bar d}\}=i\omega \delta_{a\bar d}I_{c\bar b}-
i\omega \delta_{c\bar b}I_{a\bar d}.
\label{fradkin}\ee
In the $SU(2|1)$ supersymmetric extension of this system,  the Hamiltonian, dynamical supercharges and $R$-charges are determined by those of the  two-dimensional isotropic oscillator
\be
\mathcal{H}=\sum_{a=1}^N \mathcal{H}_a,\quad  \Theta^\alpha=\sum_{a=1}^N\Theta^{a\alpha},\qquad \mathcal{R}^{\alpha}_{\beta}=\sum_{a=1}^N \mathcal{R}^{a\alpha}_{\beta}\,,
\ee
with
\be
\mathcal{H}_a=\pi_a\bar\pi_a+\omega^2z^a\bar z^a+\frac B2 \eta^{a \alpha}\bar \eta^a_\alpha
, \quad \Theta^{a\alpha}=\pi_a\eta^{a\alpha}+i \omega z^a{\varepsilon^{\alpha \beta}}\bar\eta^a_\beta, \quad \mathcal{R}^{a\alpha}_\beta=\eta^{a\alpha}\bar\eta^a_\beta -\frac12\delta^\alpha_\beta i \eta^{a\gamma}\bar\eta^a_\gamma.
\label{sho}\ee
All constants of motion of the bosonic Hamiltonian become those of the supersymmetrized   one, since all these quantities are just  sums of bosonic and fermionic parts.
Moreover, in the supersymmetric system there appear additional symmetry generators acting on the fermionic variables only.
Thus, the system with the Hamiltonian \eqref{sho} inherits kinematical  $SU(N)$ symmetries of the bosonic sector  \eqref{suN}, hidden symmetries generated
by the Fradkin tensor \eqref{fradkin}, and reveals an additional $U(N)$ symmetry realized in the fermionic sector:
\be
 \mathcal{R}_{a\bar b}=\sum_{\alpha}\eta^{b\alpha}\bar\eta^a_\alpha:
 \qquad \{\mathcal{R}_{a\bar b}, \mathcal{R}_{c\bar d}\}=i\delta_{a\bar d}\mathcal{R}_{c\bar b} -i \delta_{c\bar b }\mathcal{R}_{a\bar d}\,.
\ee
Now we turn  to considering  less trivial example of $SU(2|1)$ supersymmetric K\"ahler oscillator with hidden symmetries.

\subsubsection*{ $\mathbb{C}^N$-Smorodinsky-Winternitz system}
The  $\mathbb{C}^N$-Smorodinsky-Winternitz system is defined by the Hamiltonian \cite{shmavonyan}
\be
{H}_{SW}=\sum_{a=1}^N{I}_a, \qquad
{I}_a= \pi_a\bar{\pi}_a+|\omega|^2 z^a\bar{z}^a+\frac{|g_a|^2}{z^a\bar z^a}.
\label{CSW}\ee
It has  $N$ manifest $U(1)$ symmetries  $z^a\to {\rm e}^{i\kappa} $,  with the generators
 $J_{a\bar a}$, and the hidden symmetries spanned  by the above generators $I_{a}$,  as well as by the following ones (the so-called Uhlenbeck tensor)
\be
I_{ab}=J_{a\bar b}J_{b\bar a}-\frac12 J_{a\bar a}J_{b\bar b}+ \frac{|g_a|^2z^b\bar{z}^b}{z^a\bar{z}^a}+\frac{|g_b|^2z^a\bar{z}^a}{z^b\bar{z}^b},\; :\quad \{I_{ab}, {H}_{SW}\}=0\,,
\label{ISW}\ee
where  $J_{a\bar b}$ are $u(N)$ generators defined in \eqref{suN}.

 This system can be identified as a K\"ahler oscillator with the following K\"ahler potential
\be
{K=z\bar z+\frac{ g_a}{\omega}\log{z^a}+\frac{ {\bar g}_a}{\bar\omega}\log {{\bar z}^a},\qquad {\bf {\rm arg }}\; \omega={\bf {\rm arg}}\sum_{a=1}^N g_a+\pi/2.}
\ee
Its $SU(2|1)$ supersymmetric extension is found to be associated with the Hamiltonian
\be
\mathcal{H}_{SW}=\sum_{a=1}^N\mathcal{I}_a, \qquad
 \mathcal{I}_a= \pi_a\bar{\pi}_a+|\omega|^2 z^a\bar{z}^a+\frac{|g_a|^2}{z^a\bar z^a} +\frac{g_a}{2}\frac{\eta^{a\alpha}\eta^{a}
_{\alpha}}{z^az^a}+
 \frac{\bar g_a}{2}\frac{\bar\eta^a_{\alpha}\bar\eta^{a\alpha}}{{\bar z}^a\bar z^a}+\frac{B}{2}\eta^{a\alpha}\bar\eta^a_\alpha,
\label{SWsuper}\ee
and the supercharges
\be
    {\Theta^{a\alpha}=\pi_a\eta^{a\alpha}+i \omega\varepsilon^{\alpha \beta}\bar\eta^a_\beta \left(z^a+\frac{g_a}{\omega z^a}\right)}.
\ee
Clearly, the generators $\mathcal{I}_a$ commute with each other, and so they are the constants of motion of the supersymmetric $\mathbb{C}^{N}$-Smorodinsky-Winternitz system.
This supersymmetric system possesses $N$ manifest $U(1)$ symmetries  $z^a\to {\rm e}^{i\kappa}z^a, \;\eta^a_\alpha\to {\rm e}^{i\kappa}\eta^a_\alpha\,$, with the generators
\be
\mathcal{J}_{a\bar a}=J_{a\bar a}+\eta^{a\alpha}\bar\eta^a_\alpha\; :\{\mathcal{J}_{a\bar a}, \mathcal{J}_{b\bar b}\}=\{\mathcal{J}_{a\bar a}, \mathcal{I}_{b}\}=0\,.
\ee
The extensions  of the hidden symmetry generators $I_a, I_{ab}$ are given, respectively,  by the generators $\mathcal{I}_a$ defined  in \eqref{SWsuper}
and by the following ones
\be
\mathcal{I}_{ab}=I_{ab}+\frac{g_a}{2}\frac{z^b \bar z^b}{z^a z^a}\eta^{a\alpha} \eta^a_\alpha
+\frac{\bar g_a}{2}\frac{z^b \bar z^b}{\bar z^a \bar z^a}\bar \eta^{a}_{\alpha} \bar \eta^{a\alpha}
+\frac{g_b}{2}\frac{z^a \bar z^a}{z^b z^b}\eta^{b\alpha} \eta^b_\alpha
+\frac{\bar g_b}{2}\frac{z^a \bar z^a}{\bar z^b \bar z^b}\bar \eta^{b}_{\alpha} \bar \eta^{b\alpha}\; : \{\mathcal{I}_{ab},\mathcal{H}_{SW}\}=0\,.
\label{ISW1}\ee
Thus $SU(2|1)$ supersymmetric extension of $\mathbb{C}^{N}$-Smorodinsky-Winternitz system inherits all its hidden symmetries.\\

The conclusion is that the ``K\"ahler superoscillator approach'' yields the well defined superextensions of both the isotropic oscillator and the Smorodinsky-Winternitz
system on  $\mathbb{C}^N$.

\subsection{Complex projective spaces}
In this Section we will deal with superintegrable  systems on complex projective spaces $\mathbb{CP}^N$ which
are specified by the presence of constant magnetic field and belong to the class of the K\"ahler oscillator models.

Consider  the complex projective space equipped with  $su(N+1)$-invariant    Fubini-Study metrics
  \begin{equation}
  g_{a\bar b}dz^ad{\bar z}^b,\quad{\rm with} \quad g_{a\bar b} = \frac{\log(1+z\bar z)}{\partial z^a\partial {\bar z}^b}=
  \frac{\delta_{a\bar b}}{1+z \bar z}-\frac{{\bar z}^a z^b}{(1+z \bar z)^2 }\,.
 \end{equation}
The  inverse  metrics,  non-zero Christoffel symbols and  Riemann tensor  are defined by the expressions
\be
g^{ \bar a b}=(1+z\bar z)(\delta^{\bar a b}+  {\bar z}^a z^b),\quad \Gamma^{a}_{bc}=-\frac{\delta^a_b\bar z^c+\delta^a_c \bar z^b}{1+z\bar z}\,.
\quad
R_{a\bar b c \bar d}=g_{a\bar b}g_{c\bar d}+g_{c\bar{b}}g_{a\bar d},
\ee
The Killing potentials of   $su(N+1)$ isometry algebra are of the form
\be
h_{a\bar b}=\frac{z^b\bar{z}^a}{1+z\bar z},\qquad h_a=\frac{{\bar z}^a}{1+z\bar z}\,.
\label{suKilling}\ee
Equipping the cotangent bundle of $\mathbb{CP}^N$  with the twisted symplectic structure \eqref{ssB} and the related Poisson brackets,
we obtain the mechanics systems involving an interaction with  a constant magnetic field.

 The  $su(N+1)$ isometry generators are given by the expressions  of the form
\be\label{suN1}
\begin{array}{c}
J_{a\bar b}=i(z^b\pi_a-\bar\pi_b\bar z^a)- B\frac{\bar z^a z^b}{1+z\bar z},\quad
J_{a}=i(\pi_a+\bar z^a(\bar z\bar\pi))-B\frac{{\bar z}^a}{1+z\bar z}:\\[4mm]
\{J_{{\bar a} b}, J_{\bar c d}\}=
i\delta_{\bar a d}J_{\bar b c}
-i\delta_{\bar c b}J_{\bar a d},\quad \{J_{a}, {\bar J}_{b}\}=i J_{a\bar b},\quad \{J_a, { J}_{b\bar c}\}=\mp i J_b\delta_{a\bar c}\,.
\end{array}\ee
Extending  these generators to this   phase superspace   as in   \eqref{sNeth}, we obtain
\be
\mathcal{J}_{a\bar b}=J_{a\bar b}+
\frac{\partial^2 {h}_{a\bar b}}{\partial z^c\partial
 {\bar z}^d}\eta^{c\alpha}\bar\eta^{d}_\alpha,\quad \mathcal{J}_{a}=J_{a}+\frac{\partial^2 h_{a}}{\partial z^c\partial
 {\bar z}^d}\eta^{c\alpha}\bar\eta^{d}_\alpha.
\label{suNsuper}\ee

With these expressions at hand we can construct superintegrable  models admitting weak $SU(2|1)$ supersymmetry.

\subsubsection*{$\mathbb{CP}^N$-oscillator}
The oscillator on a complex projective space is defined by the Hamiltonian \cite{CPosc} \footnote{ Hereafter we use the notation $z\bar z\equiv \sum_{c=1}^N z^c{\bar z}^c$, $(\pi z)=
\sum_{c=1}^N\pi_c z^c$ etc. }
\be
{ H}_{osc}=g^{\bar a b}\bar\pi_{a}\pi_{b }+|\omega|^2 z\bar z\,.
\label{hn}\ee
The constants of motion of this system   are given by the  $u(N)$-generators  $J_{a\bar b}$  \eqref{suN1} and by the analog of ``Fradkin tensor"
\be\label{fradkin1}
I_{a\bar b}={J_a {\bar J}_b} +|\omega|^2 {\bar z}^a z^b\,.
\ee
This system belongs to the class of ``K\"ahler oscillators'' \eqref{Kahosc} with $K=\log(1+z\bar z)$, and hence admits  $SU(2|1)$ supersymmetric extension.
The relevant Hamiltonian and supercharges read
\be
\mathcal{H}_{osc}=g^{\bar a b}\bar\pi_a \pi_b+|\omega|^2z\bar z
-\frac 12  (g_{a\bar b}g_{c\bar d}+g_{c\bar{b}}g_{a\bar d})\eta^{a\alpha}\bar\eta^b_\alpha\eta^{c\beta}\bar\eta^d_\beta
-\frac{\omega}{2}\frac{\bar z^a \bar z^b\eta^{a\alpha}\eta^b_\alpha}{(1+z\bar z)^2}-
\frac{\bar \omega}{2}\frac{z^a  z^b\bar\eta^{a}_{\alpha}\bar\eta^{b\alpha}}{(1+z\bar z)^2} + \frac B2
{g_{a\bar b}\eta^{a\alpha}{\bar\eta}^b_\alpha },
\ee
\be
\Theta^\alpha=\pi_a \eta^{a\alpha}+ i\bar\omega \frac{z^a}{1+z \bar z} {\varepsilon^{\alpha \beta}}{\bar \eta}^{a}_{\beta},\quad \overline{\Theta}_{\alpha}=
\bar\pi_a \bar\eta^a_\alpha + i\omega\frac{\bar z^a}{1+z\bar z}{\varepsilon_{\alpha \beta}} {\eta}^{a \beta} .
\ee
This system has the manifest  $u(N)$ symmetry defined by the generators  $\mathcal{J}_{a\bar b}$: $ \{\mathcal{J}_{a\bar b},\mathcal{H}_{osc}\}=0\,$.

One could expect that the appropriate generalization of the Fradkin tensor should still have the form \eqref{fradkin1}, with  $J_a$ replaced by $\mathcal{J}_a$,
and that just this minimal modification yields constants of motion of the super-oscillator. However, one can check that it is not the case.
So, for the time being, it is an open question whether a supersymmetric counterpart of the Fradkin tensor exists.
\subsubsection*{$\mathbb{CP}^N$-Rosochatius system}
The $\mathbb{CP}^N$-Rosochatius system
is defined  by the symplectic structure \eqref{ssB} and by the Hamiltonian \cite{Ros}
\be
{ H}_{Ros}=(1+z\bar z)\left(\pi\bar\pi+(z\pi)(\bar z\bar\pi) + |\omega_0|^2+ \sum_{a=1}^N\frac{|\omega_a|^2}{z^a{\bar z}^a}\right)-\sum_{i=0}^N|\omega_i|^2.
\label{Ham}\ee
This system possesses  $N$ manifest $U(1)$ symmetries with the  generators $J_{a\bar a}$  defined in  \eqref{suN1}, as well as symmetries generated by the
 second-order constants of motion
\be
I_{a}=J_{a}{\bar J}_{\bar a} +\omega^2_0 z^a{\bar z}^a +\frac{\omega^2_a}{{\bar z}^a z^a} ,\qquad I_{ab}=J_{a\bar b}J_{b\bar a}-\frac 12J_{a\bar a}J_{b\bar b}
+\left(\omega^2_a\frac{z^b{\bar z}^b}{z^a{\bar z}^a} +
\omega^2_b\frac{z^a{\bar z}^a}{z^b{\bar z}^b}
\right).
\label{fradkin2}\ee
The Hamiltonian \eqref{Ham} can be cast, up to a constant shift,  in the form of the  ``K\"ahler oscillator" Hamiltonian \cite{CPosc,Kahlerosc}
\be
{H}_{Ros}=g^{a\bar b}\left(\pi_a\bar\pi_b + |{\omega}|^2\partial_{a}K\partial_{\bar a}K\right)-E_{0},
\label{Kosc}\ee
where
\be
 K=\log(1+z\bar z)-\sum_{a=1}^N (\frac{\omega_a}{\omega}\log z^a +\frac{{\bar \omega}_a}{\bar\omega}\log {\bar z}^a),\quad \omega={\sum_{i=0}^N\omega_i},\quad E_{0}=|\sum_{i=0}^N\omega_i|^2-\sum_{i=0}^N|\omega_i|^2.
\label{suprel}\ee
Thus this system  admits  $SU(2|1)$ supersymmetric extension given by the following Hamiltonian and supercharges
\bea
\mathcal{H}_{Ros}&=&H_{Ros}-\frac 12  (g_{a\bar b}g_{c\bar d}+g_{c\bar{b}}g_{a\bar d})\eta^{a\alpha}\bar\eta^b_\alpha\eta^{c\beta}\bar\eta^d_\beta -
\Bigg(\frac{\omega\bar z^a \bar z^b}{1+z\bar z}-\frac{ \omega_a\bar z^b}{z^a}-\frac{ \omega_b\bar z^a}{z^b}\Bigg)\frac{\eta^{a\alpha}\eta^b_\alpha}{2(1+z\bar z)}
\nonumber\\
&&-\Bigg(\frac{ \bar\omega z^a  z^b}{1+z\bar z}-\frac{ \bar \omega_a z^b}{\bar z^a}-\frac{ \bar \omega_b z^a}{\bar z^b}\Bigg)\frac{\bar\eta^{a}_{\alpha}\bar\eta^{b\alpha}}{2(1+z\bar z)}+\frac B2
 g_{a\bar b}\eta^{a\alpha}{\bar\eta}^b_\alpha ,\\
\Theta^\alpha&=&\pi_a \eta^{a\alpha}+ i\Big(\bar\omega \frac{z^a}{1+z \bar z}-\frac{\bar \omega_a}{\bar z^a}\Big){\varepsilon^{\alpha \beta}}{\bar \eta}^{a}_{\beta}.
\eea
They are easily checked to constitute the $su(2|1)$ superalgebra \eqref{oscalgsu21} ($\mathcal{H}_{Ros}\equiv \mathcal{H}_{osc}$).

It is interesting that, in contrast to $\mathbb{C}^N$-Smorodinsky-Winternitz system,
in the absence of magnetic field and under the special choice of the parameters $\omega_i\,$,
 this system admits flat $\mathcal{N}=4, d=1$ ``Poincar\'e'' supersymmetry \cite{Ros}.
The choice just mentioned is as follows
\be
B=0,\quad |\omega|=|\sum_{i=0}^N\omega_i|=0.
\ee
The second equation has the simple graphical illustration: it defines  the planar  polygon  with the edges $|\omega_a|$, and, therefore, corresponds to inequality
$|\omega_0|\leq \sum_{a=1}^N |\omega_a|\,$, where, without loss of generality, we assume that $|\omega_0|\geq |\omega_1|\geq\ldots\geq |\omega_N| $.
In this case we  arrive at the well-known $\mathcal{N}=4$ supersymmetric mechanics on K\"ahler manifold with the holomorphic prepotential
${U}(z)=\sum_{a=1}^N\omega_a\log z^a$ (see, e.g., \cite{PRDrapid}).

Finally, we note that all symmetries respected by the systems considered in this section are symmetries of the appropriate superfield Lagrangians \eqref{lagrSF}
at $B\neq 0, \omega\neq 0\,$, with $\Phi^a, \bar\Phi^b$ standing for $z^a, \bar z^b\,$.

\section{Discussion and outlook}
In this paper we presented the systematic combined Hamiltonian and superfield approach to the construction of the multi-particle models of deformed $\mathcal{N}=4,8$
supersymmetric mechanics on K\"ahler manifolds in interaction with a constant magnetic fields. The latter are introduced  via a supersymmetric  version
of minimal coupling. We applied this approach  to the  various (super)integrable models and demonstrated that
such superextensions preserve all  kinematical symmetries of the initial bosonic systems (and some  hidden symmetries in a few particular cases).
One of the basic features of our approach is that diverse isometries  are realized on the $SU(2|1)$ multiplets of the same sort,
without introducing any extra multiplet. This is a crucial difference of our approach from the models of Refs. \cite{BKS11}, \cite{BKKS11}, \cite{BKKS13} in which similar
isometries were realized within the standard ${\cal N}=4$ supersymmetric mechanics at cost of introducing extra degrees of freedom (coming back
to the spin variables introduced in \cite{SpinVar}) \footnote{Applications of the spin variables in the models of $SU(2|1)$ mechanics were considered, e.g., in \cite{FedIva}.}.

The next obvious  task is  the study of the {\sl quantum mechanical} properties (spectra, etc) of the  $SU(2|1)$ supersymmetric
Landau problem on $\mathbb{CP}^N$, as well as of the $SU(2|1)$  supersymmetric oscillator-like models on $\mathbb{C}^N$ and $\mathbb{CP}^N$.

Some other  tasks are:
\begin{itemize}
\item Coupling, to a constant magnetic field,  of ``flat'' $\mathcal{N}=8$ supersymmetric mechanics with non-zero potential on special K\"ahler manifolds suggested in \cite{knsh} and
studying the new deformed $\mathcal{N}=8$ mechanics models obtained in this way;
\item The  construction of the deformed supersymmetric
extensions of the Landau problem on quaternionic manifolds and, in
particular, on quaternionic projective spaces $\mathbb{HP}^N$,
     having in mind their relevance to the so-called high-dimensional Hall effect \cite{4hall};
  \item The construction  of the $\mathbb{HP}^N$-Rosochatius  system and studying the symmetry properties of it  and of the $\mathbb{HP}^N$-oscillator's \cite{HPN},
  as well as of their  supersymmetric extensions.

  \item Introducing the notion of quaternionic  oscillator, by analogy with the K\"ahler one, and the study of its  possible deformed supersymmetric extensions.
\end{itemize}
We plan to address this circle of  problems in a not distant future.

\acknowledgements
The authors acknowledge
 a partial support from the RFBR grant, project No. 18-02-01046 (E.I. and S.S.), the grants of Armenian Committee of Science 18RF-002 and 18T-1C106 (A.N. and H.S.), and  the Regional Doctoral Program on Theoretical and Experimental Particle Physics Program
sponsored by VolkswagenStiftung (H.S.).
 The work of A.N. and H.S. was fulfilled within the ICTP Affiliated Center Program AF-04  and ICTP Network project NT-04. E.I. thanks Sergey Fedoruk for a valuable discussion.
 S.S. thanks the Directorate of YerPhI for the kind hospitality in Yerevan extended to him within the Ter-Antonyan-Smorodinsky Program at the last stage of this work.

\appendix
\section{Isometries of special K\"ahler structure in the local coordinates}
In this Appendix we formulate the conditions \eqref{fabc} defining the isometries of {\sl special K\"ahler structure} in the local coordinate frame,
in which  the K\"ahler metric and the tensor $f_{abc}(z)$ take the  form  \eqref{localK}.
The equation \eqref{fabc}
 expresses, in the special coordinate frame,  via Seiberg-Witten prepotential ${\cal F}(z)$ as follows
\bea
    3\partial_{(a}V^{d}_{\mu}\,\partial_{b}\partial_{c)}\partial_{d}{\cal F}+V^{d}_{\mu}\,\partial_{a}\partial_{b}\partial_{c}\partial_{d}{\cal F}=0\,,\label{Fabc}
\eea
with  $V^a_\mu, {\bar V}^{\bar a}_\mu$ being the components of the holomorphic  Hamiltonian vector field \eqref{iso}.

To extract the necessary corollaries of this equation, we first act by the derivative $\partial_{a}$ on \eqref{TranK}, where the K\"ahler potential is defined by \eqref{localK}.
Step by step it yields
\bea
    &&\partial_{a}\partial_{b}\partial_{\bar c}\left[\left(V^d_{\mu}\partial_d +V^{\bar d}_{\mu}\partial_{\bar d}\right)\left(\bar{z}^e\partial_{e}{\cal F}+z^e\partial_{\bar{e}}\bar{\cal F}\right)\right]=0 \quad\Rightarrow\nn
    &&\partial_{a}\partial_{b}\partial_{\bar c}\left[V^d_{\mu}\left(\bar{z}^e\partial_d\partial_{e}{\cal F}+\partial_{\bar{d}}\bar{\cal F}\right)+V^{\bar d}_{\mu}\left(\partial_{d}{\cal F}+z^e\partial_{\bar d}\partial_{\bar{e}}\bar{\cal F}\right)\right]=0 \quad\Rightarrow\nn
    &&\partial_{a}\partial_{\bar c}\left[\partial_{b}V^{d}_{\mu}\,\partial_{\bar d}\bar{\cal F}+\bar{z}^e\partial_{b}\left(V^{d}_{\mu}\,\partial_{d}\partial_{e}{\cal F}\right)+V^{\bar d}_{\mu}\left(\partial_{\bar d}\partial_{\bar b}\bar{\cal F}+\partial_{d}\partial_{b}{\cal F}\right)\right]=0 \quad\Rightarrow\nn
    &&\partial_{\bar c}\left[V^{\bar d}_{\mu}\,\partial_{a}g_{b\bar{d}}+\partial_{a}\partial_{b}V^{d}_{\mu}\,\partial_{\bar d}\bar{\cal F}+\bar{z}^e\partial_{a}\partial_{b}\left(V^{d}_{\mu}\,\partial_{d}\partial_{e}{\cal F}\right)\right]=0 \quad\Rightarrow\nn
    &&\partial_{\bar c}V^{\bar d}_{\mu}\,\partial_{a}\partial_{b}\partial_{d}{\cal F}+\partial_{a}\partial_{b}V^{d}_{\mu}\,g_{d\bar{c}}+\partial_{a}V^{d}_{\mu}\,\partial_{d}\partial_{c}\partial_{b}{\cal F}+\partial_{b}V^{d}_{\mu}\,\partial_{d}\partial_{c}\partial_{a}{\cal F}+V^{d}_{\mu}\,\partial_{d}\partial_{c}\partial_{a}\partial_{b}{\cal F}=0\quad\Rightarrow\nn
    &&3\partial_{(a}V^{d}_{\mu}\,\partial_{b}\partial_{c)}\partial_{d}{\cal F}-\partial_{a}\partial_{b}\partial_{d}{\cal F}\left(\partial_{c}V^{d}_{\mu}-\partial_{\bar c}V^{\bar d}_{\mu}\right)+V^{d}_{\mu}\,\partial_{a}\partial_{b}\partial_{c}\partial_{d}{\cal F}+g_{d\bar{c}}\,\partial_{a}\partial_{b}V^{d}_{\mu}=0\,.
\eea
Using the last condition, we become able to rewrite \eqref{Fabc} as
\bea
    g_{d\bar{c}}\,\partial_{a}\partial_{b}V^{d}_{\mu}-\partial_{a}\partial_{b}\partial_{d}{\cal F}\left(\partial_{c}V^{d}_{\mu}-\partial_{\bar c}V^{\bar d}_{\mu}\right)=0\,.\label{Fabc2}
\eea
Next, taking $\partial_{\bar{e}}$ derivative of this relation, we obtain
\bea
    \partial_{\bar e}\partial_{\bar d}\partial_{\bar c}\bar{\cal F}\,\partial_{a}\partial_{b}V^{d}_{\mu}=-\,\partial_{a}\partial_{b}\partial_{d}{\cal F}\,\partial_{\bar{e}}\partial_{\bar c}V^{\bar d}_{\mu}\,.
\eea
The left- and right-hand sides of this relation are products of holomorphic and antiholomorphic functions. Obviously, the factors of the same holomorphicity should be equal, which yields
\bea
    \partial_{a}\partial_{b}V^{c}_{\mu}=iC^{cd}_\mu\partial_{a}\partial_{b}\partial_{d}{\cal F}, \quad C^{cd}_\mu={\bar C}^{dc}_\mu\,,
     \label{Product}
\eea
where $C^{cd}_\mu$ are some complex constant parameters.

Taking also into account \eqref{Fabc2}, the solution of \eqref{Product} can be written as
\bea
    V^{d}_{\mu} = i C^{de}_\mu\partial_{e}{\cal F}+ \beta^d_{\mu\,a}z^{a}+\alpha^d_{\mu},\quad
    V^{\bar d}_{\mu}= - i C^{de}_\mu\partial_{e}{\cal F}+ \beta^d_{\mu\,a}\bar{z}^{a}+\bar{\alpha}^d_{\mu},\label{V}
\eea
where $\beta^d_{\mu\,a}$ and $\alpha^d_{\mu}$ are, respectively, real and complex constant parameters. From \eqref{Fabc2} and \eqref{Product},
it follows that $C^{cd}_\mu$ is a symmetric real matrix, $C^{cd}_\mu=C^{dc}_\mu$\,.

The variation of ${\cal F}$ is then equal to
\bea
    \delta_{\mu}{\cal F}\equiv V^d_{\mu}\partial_{d}{\cal F}=\left(i C^{de}_\mu\partial_{e}{\cal F}+ \beta^d_{\mu\,a}z^{a}+\alpha^d_{\mu}\right)\partial_{d}{\cal F} .\label{VarF}
\eea
Inserting this solution in \eqref{Fabc} yields the condition
\bea
    \partial_{a}\partial_{b}\partial_{c}\left(\delta_{\mu}{\cal F}\right)=0\,,
\eea
having the obvious general solution
\bea
    \delta_{\mu}\mathcal{F}= c_{\mu} + c_{a\,\mu}z^a + c_{ab\,\mu}z^az^b,\label{deltaF}
\eea
where $c_{\mu}$, $c_{a\,\mu}$ and $c_{ab\,\mu}$ are complex parameters.

Next we insert the solution \eqref{V} in the Killing equation \eqref{Killing} (b), with the metric defined by \eqref{local}, and derive the additional condition on $\delta_{{\mu}}{\cal F}$:
\bea
    \partial_{a}\partial_{b}\left(\delta_{\mu}{\cal F}\right)+\partial_{\bar a}\partial_{\bar b}\left(\delta_{{\mu}}\bar{\cal F}\right)=0\,.\label{ddF}
\eea
This equation amounts to the reality condition $\overline{\left(c_{ab\,\mu}\right)}=-\,c_{ab\,\mu}$\,.

The superfield transformations \eqref{tr_F} have precisely the form of the general isometry $\delta_{\mu}{\cal F}$, with the complex coordinates $z^a, \bar{z}^a$
being replaced by the chiral $SU(4|1)$ superfields $\Phi^a$ and their anti-chiral counterparts.


\begin{thebibliography}{99}


\bibitem{GK}
L.E.~Gendenshtein, I.V.~Krive, \emph{``Supersymmetry In Quantum Mechanics"},
Sov.\ Phys.\ Usp.\ {\bf 28} (1985) 645.
\bibitem{LL}
L.D.~Landau, E.M.~Lifshitz,
{\sl Quantum Mechanics}, 4th edition, Nauka Publ., Moscow, 1973.


\bibitem{hall}
F.D.M.~Haldane,
``Fractional Quantization Of The Hall Effect: A Hierarchy Of Incompressible
Quantum Fluid States,''
Phys.\ Rev.\ Lett.\ {\bf 51} (1983) 605;

R.B.~Laughlin,
``Anomalous quantum Hall effect: An incompressible quantum fluid with
fractionally charged excitations,''
Phys.\ Rev.\ Lett.\  {\bf 50} (1983) 1395.

\bibitem{cpn}
D.~Karabali, V.P.~Nair,
``Quantum Hall effect in higher dimensions,''
Nucl.\ Phys.\  B {\bf 641} (2002) 533
[arXiv:hep-th/0203264].


\bibitem{hasebe}
K.~Hasebe,
\emph{Supersymmetric quantum Hall effect on fuzzy supersphere,}
Phys.\ Rev.\ Lett.\  {\bf 94} (2005) 206802
[arXiv:hep-th/0411137];
\emph{Hyperbolic SUSY Quantum Hall Effect,}
Phys.\ Rev.\  D {\bf 78} (2008) 125024
[arXiv:0809.4885 [hep-th]].

\bibitem{nlin}
S.~Bellucci, A.~Beylin, S.~Krivonos, A.~Nersessian, E.~Orazi,
\emph{``N = 4 supersymmetric mechanics with nonlinear chiral supermultiplet,''}
Phys.\ Lett.\  B {\bf 616} (2005) 228
[arXiv:hep-th/0503244];

   S.~Bellucci and A.~Nersessian,
  \emph{``Nonlinear chiral supermultiplet: Freedom in the fermion-boson coupling,''}
  Phys.\ Rev.\ D {\bf 73} (2006) 107701

  [hep-th/0512165].



 \bibitem{CPosc}
  S.~Bellucci and A.~Nersessian, \emph{(Super)oscillator on $\mathbb{CP}^N$ and constant magnetic field,}
  Phys.\ Rev.\ D {\bf 67}, 065013 (2003)
  Erratum: [Phys.\ Rev.\ D {\bf 71}, 089901 (2005)]
  [hep-th/0211070].

\bibitem{smilga}
  A.~V.~Smilga,
 \emph{Weak supersymmetry,}
  Phys.\ Lett.\ B {\bf 585} (2004) 173
  [hep-th/0311023].


\bibitem{Kahlerosc}
  S.~Bellucci and A.~Nersessian, \emph{Supersymmetric Kahler oscillator in a constant magnetic field,}
{\sl Proc of 5th International Workshop on Supersymmetries and Quantum Symmetries, Dubna, Russia, July 24 - 29, 2003, Ed. E.Ivanov and A Pashnev}, pp.379-483, JINR Publ., Dubna  [hep-th/0401232].

\bibitem{ivanovsidorov}
 E.~Ivanov, S.~Sidorov,
 \emph{Deformed Supersymmetric Mechanics,}
  Class.\ Quant.\ Grav.\  {\bf 31} (2014) 075013
  [arXiv:1307.7690 [hep-th]];
    E.~Ivanov, S.~Sidorov, F.~Toppan,
\emph{Superconformal mechanics in $SU(2|1)$ superspace,}
  Phys.\ Rev.\ D {\bf 91}(2015)085032
  [arXiv:1501.05622 [hep-th]].

   E.~Ivanov, O.~Lechtenfeld, S.~Sidorov,
  \emph{SU$(2|2)$ supersymmetric mechanics,}
  JHEP {\bf 1611}(2016)031
  [arXiv:1609.00490 [hep-th]].
\bibitem{ISKahler}
  E.~Ivanov and S.~Sidorov,
\emph{Super K\"ahler oscillator from $SU(2|1)$ superspace,}
  J.\ Phys.\ A {\bf 47} (2014) 292002
  [arXiv:1312.6821 [hep-th]].
\bibitem{kor}
S.T.~Hong, J.~Lee, T.H.~Lee and P.~Oh,
\emph{A complete solution of a constrained system: SUSY monopole quantum
mechanics,}
JHEP {\bf 0602} (2006) 036
[arXiv:hep-th/0511275];


\emph{N=4 supersymmetric quantum mechanics with magnetic monopole,}
  Phys.\ Lett.\ B {\bf 628} (2005) 165
  [hep-th/0507194].


\bibitem{ILS19rev} E.~Ivanov, O.~Lechtenfeld and S.~Sidorov,
 \emph{``Deformed $N$= 8 Supersymmetric Mechanics,''}
  Symmetry {\bf 11} (2019) no.2,  135.
  doi:10.3390/sym11020135

\bibitem{sca}
  G.~Festuccia, N.~Seiberg,
\emph{Rigid Supersymmetric Theories in Curved Superspace,}
  JHEP {\bf 1106}(2011)114[arXiv:1105.0689 [hep-th]];

  T.~T.~Dumitrescu, G.~Festuccia, N.~Seiberg,
\emph{Exploring Curved Superspace}
  JHEP {\bf 1208}(2012)141[arXiv:1205.1115 [hep-th]].

 \bibitem{fre}
  P.~Fre,
  \emph{``Lectures on special Kahler geometry and electric - magnetic duality rotations,''}
  Nucl.\ Phys.\ Proc.\ Suppl.\  {\bf 45BC} (1996) 59
  [hep-th/9512043].

\bibitem{shmavonyan}
  H.~Shmavonyan,
\emph{$\mathbb{C}^N$-Smorodinsky–Winternitz system in a constant magnetic field,}
  Phys.\ Lett.\ A {\bf 383} (2019) 1223
  [arXiv:1804.03721 [hep-th]].



  \bibitem{quantCPosc}
  S.~Bellucci, A.~Nersessian and A.~Yeranyan,
 \emph{Quantum oscillator on CP**n in a constant magnetic field,}
  Phys.\ Rev.\ D {\bf 70}, 085013 (2004)
  [hep-th/0406184].


 \bibitem{Ros}
  E.~Ivanov, A.~Nersessian and H.~Shmavonyan,
 \emph{$\mathbb{CP}^N$-Rosochatius system, superintegrability, supersymmetry,}
  Phys.\ Rev.\ D {\bf 99} (2019) 085007
  [arXiv:1812.00930[hep-th]].





\bibitem{ivanov}
E.~Ivanov, L.~Mezincescu, P.K.~Townsend,
``Planar super-Landau models,''
JHEP {\bf 0601} (2006) 143
[arXiv:hep-th/0510019];\\
A.~Beylin, T.L.~Curtright, E.~Ivanov, L.~Mezincescu, P.K.~Townsend,
``Unitary Spherical Super-Landau Models,''
JHEP {\bf 0810} (2008) 069
[arXiv:0806.4716 [hep-th]].

\bibitem{n8}
  S.~Bellucci, S.~Krivonos and A.~Nersessian,
 \emph{N=8 supersymmetric mechanics on special Kahler manifolds}
  Phys.\ Lett.\ B {\bf 605} (2005) 181
  [hep-th/0410029].

 \bibitem{SW}
  N.~Seiberg and E.~Witten,
\emph{Electric - magnetic duality, monopole condensation, and confinement in N=2 supersymmetric Yang-Mills theory,}
  Nucl.\ Phys.\ B {\bf 426} (1994) 19
   Erratum: [Nucl.\ Phys.\ B {\bf 430} (1994) 485]
  [hep-th/9407087].

 \bibitem{PRDrapid}
  S.~Bellucci and A.~Nersessian,
  \emph{A Note on N=4 supersymmetric mechanics on Kahler manifolds,}
  Phys.\ Rev.\ D {\bf 64} (2001) 021702
  [hep-th/0101065].
\bibitem{BKS11}

S.~Bellucci, S.~Krivonos and A.~Sutulin, \emph{$CP^n$ supersymmetric mechanics in $U(n)$ background gauge fields}, Phys. Rev. D {\bf 84} (2011) 065033,
[arXiv:1106.2435 [hep-th]].

\bibitem{BKKS11}

S.~Bellucci, N.~Kozyrev, S.~Krivonos, and A.~Sutulin, \emph{${\cal N} =4$ chiral supermultiplet interacting with a magnetic field}, Phys. Rev. D {\bf 85} (2012) 065024,
[arXiv:1112.0763 [hep-th]].

\bibitem{BKKS13}

S.~Bellucci, N.~Kozyrev, S.~Krivonos, and A.~Sutulin, \emph{Symmetries of ${\cal N} =4$ supersymmetric $CP^n$ mechanics}, J. Phys. A {\bf 46} (2013) 275305,
[arXiv:1206.0175 [hep-th]].

\bibitem{SpinVar}

S.~Fedoruk, E.~Ivanov, O.~Lechtenfeld, \emph{Supersymmetric Calogero models by gauging}, Phys. Rev. D {\bf 79} (2009) 105015,
[arXiv:0812.4276 [hep-th]].


\bibitem{SU41}
   E.~Ivanov, O.~Lechtenfeld, S.~Sidorov,
  \emph{Deformed N=8 mechanics of (8,8,0) multiplets,}
  JHEP {\bf 1808}(2016) 193, [arXiv:1807.11804 [hep-th]].

\bibitem{dWVP}
B.~de Wit, A.~Van Proeyen, \emph{Special geometry and symplectic transformations}, Nucl. Phys. Proc. Suppl. {\bf 45BC} (1996) 196-206, [arXiv: hep-th/9510186].

\bibitem{FedIva}
S.~Fedoruk, E.~Ivanov, \emph{Gauged spinning models with deformed supersymmetry}, JHEP {\bf 1611} (2016) 103, [arXiv:1610.04202 [hep-th]].
\bibitem{knsh}
  S.~Krivonos, A.~Nersessian and H.~Shmavonyan,
\emph{``Geometry and integrability in $\mathcal{N}=8$ supersymmetric mechanics,''}
  arXiv:1908.06490 [hep-th].


\bibitem{4hall}
S.C.~Zhang, J.P.~Hu,
\emph{``A Four Dimensional Generalization of the Quantum Hall Effect,''}
Science {\bf 294} (2001) 823

  \bibitem{HPN}
  S.~Bellucci, S.~Krivonos, A.~Nersessian and V.~Yeghikyan,
\emph{Isospin particle systems on quaternionic projective spaces,}
  Phys.\ Rev.\ D {\bf 87} (2013) no.4,  045005
  [arXiv:1212.1663 [hep-th]].

\end{thebibliography}
\end{document}